\documentclass[12pt]{article}

\usepackage{chet}
\usepackage{textcomp}

\usepackage{array, caption, floatrow, tabularx, makecell, booktabs}%
\usepackage{amsfonts}
\usepackage{amssymb}
\usepackage{amsthm}
\usepackage{amsbsy}
\usepackage{bm}
\usepackage{mathrsfs}
\usepackage{slashed}
\usepackage{graphicx}
\usepackage{tikz-cd}
\usepackage{fancyhdr}
\usepackage{tikz}
\usepackage{comment}
\usepackage{xcolor}
\usepackage{capt-of}
\usepackage{chngcntr}
\usepackage[font=footnotesize]{caption}

\newcommand{\thetab}{\bar{\theta}}

\newcommand{\sigmabar}{\bar{\sigma}}

\newcommand{\dbar}{\bar{d}}

\newcommand{\pd}{\partial}
\newcommand{\al}{\alpha}
\newcommand{\bt}{\beta}
\newcommand{\ad}{\dot{\alpha} }
\newcommand{\bd}{\dot{\beta}}

\newcommand{\bra}{\langle}
\newcommand{\ket}{\rangle}

\preprint{ }

\title{Higher-Spin States of the Superstring in an Electromagnetic Background}

\author{Karim Benakli$^{\ast}$\email{kbenakli@lpthe.jussieu.fr},
Nathan Berkovits$^{\dagger}$\email{nathan.berkovits@unesp.br}, Cassiano A. Daniel$^{\dagger}$\email{c.daniel@unesp.br} and Matheus Lize$^{\dagger}$\email{matheuslize@icloud.com}}

\affiliation{$^{\ast}$Laboratoire de Physique Th\'{e}orique et 
Hautes Energies - LPTHE\\ UMR 7589,
Sorbonne Universit\'e et CNRS, 4 place Jussieu, 75252 Paris Cedex 05, France \\
$^{\dagger}$ICTP South American Institute for Fundamental Research \\
Instituto de F\'{i}sica Te\'{o}rica, Universidade Estadual Paulista \\
Rua Dr. Bento Teobaldo Ferraz 271, 01140-070, S\~{a}o Paulo - SP, Brasil}

\abstract{Constructing a consistent four-dimensional Lagrangian for charged massive higher-spin fields propagating in an electromagnetic background is an open problem. In 1989, Argyres and Nappi used bosonic open string field theory to construct a Lagrangian for charged massive spin-2  fields in a constant electromagnetic background. In this paper, we use the four-dimensional hybrid formalism for open superstring field theory to construct a supersymmetric Lagrangian for charged massive spin-2 and spin-3/2 fields in a constant electromagnetic background. The hybrid formalism has the advantage over the RNS formalism of manifest $\mathcal{N}=1$ d=4 spacetime supersymmetry so that the spin-2 and spin-3/2 fields are combined into a single superfield and there is no need for picture-changing or spin fields.}

\date{October 2021}

\begin{document}

\maketitle

\toc
 
\newpage
 
\newsec{Introduction}

Charged hadronic resonances of  spin-3/2 and spin-2 exist in Nature, examples are $\Delta(1232)$ and $\pi_2(1670)$, respectively. At sufficiently low energies such composite states can be considered with a good approximation as relativistic localized particles. Therefore, one expects to be able to construct an effective field theory to describe their propagation in an electromagnetic background. 
For free fields, a systematic way to build such Lagrangians was given by Singh and Hagen in 1974 \cite{Singh:1974qz,Singh:1974rc}, but the original problems of interaction with an electromagnetic background in four dimensions and writing down the effective Lagrangian for the above-mentioned resonances are still unsolved.

For the case of spin-2, progress has been achieved. Starting with a free linearized spin-2 Lagrangian and adding minimal coupling to an electromagnetic gauge field results in the pathological propagation of a wrong number of degrees of freedom. But
at linear order in the electromagnetic field strength, there is a unique cure to this problem which rests on adding to the Lagrangian a non-minimal coupling term introduced by Federbush in \cite{Federbush:1961aa}. In fact, if one insists on a description of the propagation of just the charged spin-2 degrees of freedom in a constant electromagnetic background, the only available Lagrangian today is the Federbush one, and it was adopted for example by \cite{deRham:2014tga}. On the other hand, from the investigation of the propagation of the first massive string excitation using open bosonic string field theory, Argyres and Nappi constructed a consistent Lagrangian for a charged massive spin-2 field in $d=26$ dimensions \cite{Argyres:1989qr,Argyres:1989cu}. Under dimensional reduction,  it was  shown in \cite{Porrati:2011uu} that there is no propagating spin-1 state and one gets in four dimensions a theory of a coupled system of charged massive spin-2 and spin-0 fields.

It is curious that the expansion of the Argyres-Nappi and Federbush Lagrangians at linear order in the electromagnetic field strength do not coincide. While the latter describes the correct number of degrees of freedom for a spin-2, it predicts a gyromagnetic ratio $g=1/2$ and the corresponding equations of motion allow faster-than-light propagation \cite{Kobayashi:1987rt,Kobayashi:1978mv}. Instead, the Argyres-Nappi Lagrangian predicts the desired \cite{Ferrara:1992yc} gyromagnetic $g=2$ value needed for causality \cite{Buchbinder:2012iz,Henneaux:2013vca}, but in order to describe the correct number of degrees of freedom one needs to keep the whole nonlinear dependence in the electromagnetic field \cite{Porrati:2010hm}. 

Less progress has been made for the spin-$3/2$ case. The coupling is required to be a nonlinear function of the electromagnetic field strength  \cite{Deser:2000dz}, and the construction of a consistent Fierz-Pauli system of equations is a priori possible \cite{Porrati:2009bs}. A nonlinear function was found in \cite{Porrati:2009bs} which is defined in terms of an implicit equation that can be solved perturbatively in powers of the field strength. It is not known if other functions exist, and a tentative Lagrangian at leading order in the electromagnetic field strength, with a set of auxiliary fields,  has been proposed in \cite{Buchbinder:2015uea}, followed by a similar quadratic expansion for the case of integer spin fields  \cite{Buchbinder:2015apa}. Promoting electromagnetism and gravity to be dynamical, one can use for the spin-$3/2$ Lagrangian that of a gravitino charged under the $U(1)$ graviphoton of $\mathcal{N}=2$ supergravity. It was shown in \cite{Deser:2001dt}, generalizing the proof of causality in supergravity \cite{Deser:1977uq}, that this allows to solve the problem of faster-than-light propagation but only for spin-$3/2$ states of Planckian mass (see also \cite{Porrati:2009bs} for discussion of this point). 

String theory was originally constructed as a model to describe the hadronic resonances and the spectrum contains massive higher spin states, including massive spin-$3/2$ and spin-2 states. Note that open strings come with charges $q_0$ and $q_\pi$ located at the $\sigma=0$ and $\sigma=\pi$ ends, where $0\leq\sigma\leq\pi$ parametrizes the position on the open string. Therefore, fields of charge $Q$ arise from open string state configurations with $Q=q_0 + q_\pi$. As mentioned above, open bosonic string field theory was used by Argyres and Nappi to construct a consistent theory of charged massive spin-2.  

It is the purpose of this work to perform a similar construction using open superstring field theory which includes both the massive spin-$3/2$ and spin-2 states. Although one can in principle use the Ramond-Neveu-Schwarz formalism of open superstring field theory to perform these computations, we will instead use the four-dimensional hybrid formalism of open superstring field theory for two reasons. Firstly, the hybrid formalism has manifest $\mathcal{N}=1$ $d=4$ spacetime supersymmetry which will allow us to combine the spin-$3/2$ and spin-2 fields into a single $d=4$ superfield and compute the Lagrangian and equations of motion in superspace. Secondly, the hybrid formalism avoids the complicated picture-changing operators and spin fields which are necessary in the Ramond-Neveu-Schwarz formalism to describe the spin-$3/2$ states.

In section \ref{review} of this paper we give a concise review of the hybrid formalism for the superstring, stating its key properties and presenting the relevant variables in terms of the oscillator modes of free fields. We also show how to compute the equations of motion and the open superstring field theory action in this formalism. Section \ref{chargedstrings} is about charged superstrings in a constant electromagnetic background, and we start by solving the equations of motion and boundary conditions for the bosonic worldsheet fields and expressing the solution in terms of modes. The mode expansions for the fermionic worldsheet fields are then found by imposing that the left and right-moving superconformal generators are equal at the boundary. We also write down the worldsheet action for the charged superstring in the hybrid formalism and describe how the superconformal generators of the previous section are modified in the interacting case. In section \ref{spin1section} we use the machinery developed in section \ref{chargedstrings} to compute the action for the massless sector of the open superstring compactified to four dimensions in a constant electromagnetic background, and show that the results are consistent with earlier computations done for the massless sector of the bosonic string. Section \ref{actionmass} presents the computation of the $\mathcal{N}=1$ $d=4$ superspace action for the first massive compactification-independent states of the charged open superstring in a constant electromagnetic background, which on-shell describes 12 bosonic and fermionic complex degrees of freedom including a charged massive spin-$3/2$ and spin-$2$ field. In section \ref{secsix} we compute the equations of motion in Lorenz gauge for these massive interacting states and expand the answer in terms of the component fields. Finally, section \ref{conclu} contains some concluding remarks and possible future applications, and appendix \ref{spinorconventions} collects information about conventions and important equations used throughout this work.

\newsec{Review of the Hybrid Formalism}[review] 

In this section, we introduce the manifestly four-dimensional spacetime supersymmetric description of the superstring. The worldsheet action, relevant variables and their main properties are presented, as well as a formulation in terms of oscillator modes for the case without a background $U(1)$ gauge field. For further details, we refer to the original works \cite{Berkovits:1996bf} \cite{Berkovits:1994vy} \cite{Berkovits:1995ab}, by one of the authors, and \cite{Kappeli:2006fj}. Our conventions for the worldsheet theory follow \cite{Polchinski:1998rq} and we are using $\al'=\frac{1}{2}$ when the string constant is omitted. For manipulations with sigma matrices and \textit{dotted}/\textit{undotted} spinor indices we utilize the conventions in the appendix of \cite{Wess:1992cp}, for example, $x_{\al \ad} = \sigma^m_{\al \ad} x_m$, $\sigmabar^{ \ad \al}_m \sigma^m_{\bt \bd} = -2 \delta^{\al}_{\bt} \delta^{\ad}_{\bd}$, $(\psi \chi) = \psi^{\al} \chi_{\al}$, $(\bar{\psi} \bar{\chi})= \bar{\psi}_{\ad} \bar{\chi}^{\ad}$, $(\sigma^m \sigmabar^n + \sigma^n \sigmabar^m)_{\al}^{\; \bt} = -2 \eta^{mn} \delta^{\bt}_{\al} $, etc. 

\subsec{Open superstring worldsheet action and hybrid variables}

The Euclidean worldsheet action of the four-dimensional spacetime part consists of four bosons $x^m$, $m=0$ to $3$,
with two pairs of left-moving canonically 
conjugate Weyl fermions $(p^{\alpha},\theta_{\beta}) $ 
and $(\bar{p}_{\ad},\bar{\theta}^{\bd}) $ $(\al, \ad = 1,2)$ having conformal weight $(1, 0)$ each, and 
a chiral boson $\rho$. We also have the right-moving variables which will be denoted by a \textit{hat}. 

In conformal gauge, the action is given by \eqn{
S_0 =\frac{1}{2 \pi} \int d^2z \bigg\{ \frac{1}{\al'} \pd x^m \bar{\pd}x_m + p^\al \bar{\pd} \theta_\al + \bar{p}_{\ad} \bar{\pd} \bar{\theta}^{\ad} + \hat{p}^{\al} \pd \hat{\theta}_{\al} + \hat{\bar{p}}_{\ad} \pd \hat{\bar{\theta}}^{\ad} + \frac{1}{2} \pd \rho \bar{\pd} \rho + \frac{1}{2} \pd \hat{\rho} \bar{\pd} \hat{\rho} \bigg\}  + S_6, 
}[s_0]
where $\eta^{mn}= \text{diag}(-,+,+,+)$, $\pd \equiv \pd_z $ and $\bar{\pd} \equiv \partial_{\bar{z}}$. Middle alphabet letters such as $m, n, p $ will be used to denote four-dimensional spacetime indices throughout this work. 

In the strip, the Euclidean coordinates take the standard values: $0 \leq\sigma \leq \pi$ and $-\infty < \tau < \infty$, where $z = e^{-i w}$ with $w = \sigma + i \tau$. In the action, to go from the plane to the strip,  one just substitutes $d^2z = 2 d \sigma d\tau$, $\pd = \pd_w$ and $\bar{\pd} = \pd_{\bar{w}}$, where $\pd_w = \frac{1}{2} (\pd_\sigma  - i \pd_\tau)$ and $\partial_{\bar{w}}= \frac{1}{2} (\pd_\sigma + i \pd_\tau)$. As is commonly done, we will use a bar to denote complex conjugation of $z$ and $w$, but this should not be confused with $\bar{p}_{\ad}$ and $\thetab^{\ad}$ which are left-moving variables.

The worldsheet fields in the hybrid formalism are related to those in the gauge-fixed RNS description by a field redefinition \cite{Berkovits:1996bf}. The internal six dimensional matter part of the action, $S_6$, is the same as in RNS. Without loss of generality, we will suppress the right-moving fields in the rest of this paper. In the free case, they are related to the left-moving ones in the usual way by the boundary conditions, for example, $p^{\al}(z) = \hat{p}^{\al}(\bar{z})$, $\theta^{\al}(z) = \hat{\theta}^{\al}(\bar{z})$ at $\text{Im}\{z\}=0$ and $x^m$ satisfies Neumann boundary conditions.

We group the RNS matter variables for the internal directions $(x^{\mu}, \psi^{\mu})$, $\mu = 4$ to $9$, into a \textbf{3} and $\bar{\boldsymbol{3}}$ of $SU(3)$ and denote these variables by $(x^k, \bar{x}_k, \psi^k, \bar{\psi}_k)$ with indices $j, k, l$ running from $1$ to $3$. The description that we shall use corresponds to an uncompactified superstring if $x^j$ takes value on $\mathbb{R}^6$, or to a toroidally-compactified superstring if $x^j$ takes value on $T^6$. The free field OPE's in the complex plane for the four-dimensional part are \eqna{
& p^{\al}(y) \theta_{\bt}(z) \thicksim \frac{\delta^\al_\bt}{y-z}, \quad \quad \bar{p}_{\ad} (y) \bar{\theta}^{\bd} (z) \thicksim \frac{\delta^{\bd}_{\ad}}{y-z}, \\
&\rho(y) \rho(z) \thicksim \log(y-z), \quad x^m(y) x^n(z) \thicksim - \frac{\al'}{2}  \eta^{mn} (\log|y-z|^2 + \log|y-\bar{z}|^2).
}[OPEsworldsheet]

And for the internal part, we have \eqna{
&\psi^j(y) \bar{\psi}_k(z) \thicksim -\frac{\delta^j_k}{y-z}, \quad x^{j}(y) \bar{x}_k(z) \thicksim - \frac{\al'}{2} \delta^j_k (\log|y-z|^2 + \log|y-\bar{z}|^2), \\
&  H_C(y) H_C(z) \thicksim -3 \log (y-z),
}[]
where we defined $i\psi^k \bar{\psi}_k =  \pd H_C$ through bosonization.

The action is invariant under four-dimensional spacetime supersymmetry generated by \eqna{
&Q_\al = \oint (p_{\al} -  i \sqrt{\frac{2}{\al'}}  \bar{\theta}^{\ad} \pd x_{\al \ad} + \frac{1}{2} \bar{\theta}^2 \pd \theta_{\al}), \\ 
& \bar{Q}^{\ad} = \oint (\bar{p}^{\ad} - i \sqrt{\frac{2}{\al'}}\theta_{\al} \pd x^{\ad \al} + \frac{1}{2} \theta^2 \pd \bar{\theta}^{\ad}),
}[]
where $ \oint  \equiv \frac{1}{2 \pi i} \oint dz$, which satisfy the usual supersymmetry algebra \eqna{
& \{Q_{\al} , \bar{Q}_{\ad} \} = -2i \sqrt{\frac{2}{\al'}}\oint \pd x_{\al \ad}, \\
&\{Q_{\al} , Q_{\bt}\} = \{ \bar{Q}_{\ad}, \bar{Q}_{\bd} \} = 0.
}[]

Note that the four-dimensional supersymmetry generators commute with all worldsheet fields related to the compactified directions. Relevant supersymmetric combinations of the worldsheet variables are \eqna{
&d_\al = -p_\al -i \sqrt{\frac{2}{\al'}}\bar{\theta}^{\ad} \pd x_{\al \ad} + \bar{\theta}^2 \pd \theta_\al - \frac{1}{2}\theta_\al \pd \bar{\theta}^2, \\
& \bar{d}_{\ad} = -\bar{p}_{\ad} + i \sqrt{\frac{2}{\al'}} \theta^{\al} \pd x_{\al \ad} + \theta^2 \pd \bar{\theta}_{\ad} - \frac{1}{2} \bar{\theta}_{\ad} \pd \theta^2, \\
& \Pi_{\al \ad} = \sqrt{\frac{2}{\al'}}\pd x_{\al \ad} + 2i \pd \theta_{\al} \bar{\theta}_{\ad} + 2i \pd \bar{\theta}_{\ad} \theta_{\al},
}[susyvariables]
with the following OPE's \eqna{
& d_{\al}(y)\Pi_{\bt \bd}(z) \thicksim \frac{4i \epsilon_{\al \bt} \pd \bar{\theta}_{\bd}}{y-z} , \quad  \bar{d}_{\ad}(y) \Pi_{\bt \bd}(z) \thicksim \frac{-4i \epsilon_{\ad \bd} \pd \theta_{\bt}}{y-z}, \\
&d_{\al}(z)\bar{d}_{\ad}(y) \thicksim \frac{2i \Pi_{\al \ad}}{y-z} .
}[OPEsSusy]

The hermiticity conditions are defined as $(p^\al)^\dagger = -\bar{p}^{\ad}$, $(\theta^\al)^\dagger = \bar{\theta}^{\ad}$, $(\pd x_{\al \ad})^\dagger = -\pd x_{\al \ad}$, $(d_{\al})^\dagger = -\bar{d}_{\ad}$, $(\pd \rho)^\dagger = - (2 \pd \rho - \pd H_C)$ and $ (\pd H_C)^\dagger = - (3 \pd \rho -2 \pd H_C)$\foot{We do not discuss here the unusual complex conjugation properties defined for $\pd \rho$ and $\pd H_C$ and recommend \cite{Berkovits:1996bf} and \cite{Kappeli:2006fj} for an explanation, where it is referred as the $\thicksim$ conjugation. 
}. Note that $(\pd \theta^\al)^\dagger = - \pd \bar{\theta}^{\ad}$ using the standard CFT rule for a primary field $\phi$ of conformal weight $h$ on the plane, namely, $[\phi(z)]^{\dagger}= \phi^{\dagger}(\bar{z}^{-1}) \bar{z}^{-2h}$.

From the worldsheet fields, one can form the generators of a twisted small $\mathcal{N}=4$ algebra \eqna{
&T = T_4 + T_6, \quad J= J_4 + J_6, \\
&G^{\pm} = G^{\pm}_4 + G^{\pm}_6, \quad \tilde{G}^{\pm} = \tilde{G}^{\pm}_4 + \tilde{G}^{\pm}_6, \\
&J^{++} = e^{-i \rho + i H_C}, \quad J^{--}= e^{i \rho - iH_C},
}[]
where $T$ is the stress tensor and \begingroup
\allowdisplaybreaks
\begin{align*}
& T_4 = \frac{1}{4} \Pi^{\ad \al} \Pi_{\al \ad} + \pd \theta^{\al} d_{\al} + \pd \bar{\theta}_{\ad} \dbar^{\ad} + \frac{1}{2} \pd \rho \pd \rho - \frac{i}{2} \pd^2 \rho, \quad J_4= -i \pd \rho, \\ 
&T_6 = - \frac{2}{\al'}\pd \bar{x}_k \pd x^k - \pd \psi^k \bar{\psi}_k, \quad J_6= -\psi^k \bar{\psi}_k = i \pd H_C, \\
&G^+_4 = \frac{1}{2\sqrt{8}} e^{i \rho} d^2, \quad G_4^- = - \frac{1}{2\sqrt{8}} e^{-i \rho} \dbar^2, \\
&\tilde{G}^+_4 = - \frac{1}{2\sqrt{8}} e^{-2i \rho + iH_C} \dbar^2, \quad \tilde{G}^-_4 = - \frac{1}{2\sqrt{8}} e^{2i \rho -i H_C} d^2, \\
&G_6^+ = \sqrt{\frac{2}{\al'}} \partial \bar{x}_j \psi^j, \quad G_6^- = \sqrt{\frac{2}{\al'}}  \pd x^j \bar{\psi}_j,\\
&\tilde{G}^+_6 = \frac{1}{2} \sqrt{\frac{2}{\al'}}  e^{-i \rho} \epsilon_{jkl} \partial x^j \psi^k \psi^l, \quad \tilde{G}^-_6 = - \frac{1}{2} \sqrt{\frac{2}{\al'}}  e^{i \rho} \epsilon^{ijk} \pd \bar{x}_i \bar{\psi}_j \bar{\psi}_k.
\end{align*}%
\endgroup

As one can see, the small $\mathcal{N}=4$ algebra includes four supercurrents ($G^{\pm}$, $\tilde{G}^{\pm}$) and three spin-1 currents ($J$, $J^{++}$, $J^{--}$), which generate an $SU(2)$ algebra. A few OPE's these generators satisfy are \eqna{
&G_4^+(y) G_4^-(z) \thicksim -\frac{1}{(y-z)^3} + \frac{J_4}{(y-z)^2} + \frac{ T_4}{(y-z)}, \\
&G_6^+(y)G_6^-(z) \thicksim \frac{3}{(y-z)^3} + \frac{J_6}{(y-z)^2} + \frac{T_6}{(y-z)}, \\
&G^+_4(y)\tilde{G}^+_4(z) \thicksim - \frac{1}{(y-z)^2}e^{-i \rho + i H_C} + \frac{1}{(y-z)} \pd e^{-i\rho} e^{iH_C}, \\
&G_6^+(y)\tilde{G}_6^+(z) \thicksim \frac{3}{(y-z)^2} e^{-i\rho + iH_C} + \frac{1}{(y-z)} e^{-i\rho} \pd e^{iH_C}, \\
&J^{\pm \pm}(y) G^{\mp}(z) \thicksim \mp \frac{\tilde{G}^\pm}{y-z}, \\
& \Big\{ \oint  \; G^+(z), \oint  \; \tilde{G}^+(z) \Big\} = 0.
}[]

Some comments are in order. First, note that the generators $(G^{\pm}, T, J)$ form a twisted $c=6$ $\mathcal{N}=2$ superconformal field theory (SCFT). This SCFT splits in two parts: one describing the four-dimensional spacetime $(G^{\pm}_4, T_4, J_4)$ as a twisted $c=-3$ $\mathcal{N}=2$ SCFT and the other describing the six dimensional internal part ($G^{\pm}_6, T_6, J_6)$ as a  twisted $c=9$ $\mathcal{N}=2$ SCFT. The small $\mathcal{N}=4$ algebra above is then constructed by adding the currents $J^{++}$ and $J^{--}$ to form the $SU(2)$ triplet $(J, J^{++}, J^{--})$. 

One should observe that $(G^{\pm}_4, T_4, J_4)$ and $(G^{\pm}_6, T_6, J_6)$ decouple from each other, i.e., they have non-singular OPE's between them. Consequently, the six dimensional background can be replaced by any Calabi-Yau background described by an $\mathcal{N}=2$ SCFT.  Another important fact is that, in the twisted case (considered in this work), the $TT$ OPE has no conformal anomaly so one can use topological methods to compute the spectrum of correlation functions \cite{Berkovits:1994vy} without the need of introducing superconformal ghosts.

We also define the $\rho$-charge of an operator $\mathcal{O}$ as the single pole in the OPE of $J_4$ with $\mathcal{O}$, and the \textit{Calabi-Yau}-charge (CY-charge) as the single pole in the OPE of $J_6$ with $\mathcal{O}$. Properties of the generators and the hybrid variables for the twisted case are summarized in the following tables \begin{center}
\begin{minipage}{0.45\textwidth}
\centering
\begin{tabular}{ |c|c|c|c|} 
\hline
& Weight & CY-charge & $\rho$-charge  \\
\hline
$\psi^j$ & 0 & 1 & 0\\ 
\hline
$\bar{\psi}_j$ & 1 & -1 & 0\\
\hline
$e^{in \rho}$ & $\frac{-n(n+1)}{2}$ & 0 & n\\
\hline
$e^{i H_C}$ & 0 & 3 & 0\\
\hline
$e^{-i H_C}$ & 3 & -3 & 0 \\
\hline
\end{tabular}
\captionof{table}{Conformal weight, CY-charge and $\rho$-charge in the twisted case.}
\end{minipage}
\quad \quad \quad
\begin{minipage}{0.3\textwidth}
\centering
\begin{tabular}{ |c|c| } 
\hline
Generator & Weight  \\
\hline
$G^{+}, \tilde{G}^+,J$ & 1\\ 
\hline
$G^-, \tilde{G}^-, J^{--}, T $ & 2\\
\hline
$J^{++}$ & 0 \\
\hline
\end{tabular}
\captionof{table}{Conformal weight of the twisted small $\mathcal{N}=4$ generators.}
\end{minipage}
\end{center}

\subsec{Free field oscillator expansions}

In this work, we will need the description in terms of oscillator modes of the worldsheet fields. Considering first the free case, the oscillator expansions in the complex plane for the four-dimensional variables are \eqna{
&p_{\al}(z) = \sum_N \frac{p_{\al N}}{z^{N+1}}, \quad \bar{p}_{\ad}(z) = \sum_N \frac{\bar{p}_{\ad N}}{z^{N+1}}, \\
& \theta_\al(z) = \sum_N \frac{\theta_{\al N}}{z^{N}}, \quad \bar{\theta}_{\ad}= \sum_N \frac{\bar{\theta}_{\ad N}}{z^{N}}, \\
& x^m = x^m_0 - i \al'p^m \log |z|^2 + i \left(\frac{\al'}{2}\right)^{1/2} \sum_{N \in \mathbb{Z} - \{0\} } \frac{\al_N^m}{N}(z^{-N} + \bar{z}^{-N}),
}[]
where $\al_0^m = (2 \al')^{1/2} p^m$, $x^m$ satisfies Neumann boundary conditions and capital middle alphabet letters, such as $M$ and $N$, are used to denote the oscillator numbers. We also used that $\phi(z) = \sum_{ N } \frac{\phi_N}{z^{N+h}}$ for a primary field $\phi(z)$ of conformal weight $h$. 

The hermiticity properties for the modes are $(\theta^\al_N)^\dagger = \bar{\theta}^{\ad}_{-N}$, $(p^{\al}_N)^\dagger = - \bar{p}^{\ad}_{-N}$ and $(\al^m_N)^\dagger = \al^m_{-N}$, with the commutation relations \eqn{
[\al^m_M, \al^n_N] = M \delta_{M+N,0} \eta^{mn}, \quad \{ p^{\bt}_M, \theta_{\al N}\} =\delta^{\bt}_{\al} \delta_{M + N, 0}, \quad \{ \bar{p}_{\bd M}, \bar{\theta}^{\ad}_N \} = \delta^{\ad}_{\bd}  \delta_{M + N, 0},
}[]
giving the OPE's \OPEsworldsheet.

The supersymmetric variables \susyvariables in terms of the free field oscillators have the mode expansions \eqna{
&d_{\al N} = -p_{\al N} - \sum_R (\sigma^m \bar{\theta}_R)_{\al} \al_{m N-R} - \sum_{R,S}(N-2R-S) (\bar{\theta}_R \bar{\theta}_S) \theta_{\al N-R-S}, \\
&\bar{d}_{\ad N} = - \bar{p}_{\ad N} + \sum_R (\theta_R \sigma^m)_{\ad} \al_{m N-R} - \sum_{R,S} (N-2R -S) (\theta_R \theta_S) \bar{\theta}_{\ad N-R-S}, \\
& \Pi_{\al \ad N} = -i \sigma^m_{\al \ad} \al_{m N} + 2i \sum_R (N -2R) \theta_{\al R} \bar{\theta}_{\ad N-R},
}[]
where
\eqn{
d_{\al}(z) = \sum_N \frac{d_{\al N}}{z^{N+1}}, \quad \dbar_{\ad}(z) = \sum_N \frac{ \dbar_{\ad N}}{z^{N+1}}, \quad \Pi_{\al \ad}(z) = \sum_N \frac{\Pi_{\al \ad N}}{z^{N+1}}.
}

And these modes satisfy a set of commutation relations \eqna{
&[d_{\al N} ,\Pi_{\bt \bd M}] = 4i \epsilon_{\al \bt} \pd \bar{\theta}_{\bd N+M}, \quad [\bar{d}_{\ad N}, \Pi_{\bt \bd M}]= -4i \epsilon_{\ad \bd} \pd \theta_{\bt N+M}, \\
& \{d_{\al M}, \pd \theta^{\bt}_N\}= -N \delta^{\bt}_{\al} \delta_{M+N,0}, \quad \{\bar{d}^{\ad}_M, \pd \bar{\theta}_{\bd N}\}= -N \delta^{\ad}_{\bd} \delta_{M+N,0}, \\
& \{d_{\al N},\bar{d}_{\ad M}\}= 2i \Pi_{\al \ad M+N},
}[commfreeSusy]
with $\pd \theta^{\al }_N \equiv  \oint dz \; z^N\pd \theta^{\al}(z)$ and $\pd \bar{\theta}_{\ad N} \equiv \oint dz \; z^N \pd \bar{\theta}_{\ad}(z)$. 

The Virasoro generators of the four-dimensional part are defined as \eqna{
&T_4(z) = \sum_N \frac{L_N}{ z^{N + 2}} + \frac{1}{2} \pd \rho \pd \rho(z) - \frac{i}{2} \pd^2 \rho(z), \\
 &L_M = \sum_N \bigg( \frac{1}{4} \Pi^{\ad \al}_N \Pi_{\al \ad -N + M} + \pd \theta^{\al}_N d_{\al -N + M} + \pd \bar{\theta}_{\ad N} \dbar^{\ad}_{-N + M} \bigg).
}[Virasorofree]
Notice that the mode $L_0$ has no normal ordering constant. And we will write  \eqn{
d^2(z) = \sum_N \frac{d^2_N}{z^{N+2}}, \quad \dbar^2(z)= \sum_N \frac{\dbar^2_N}{z^{N+2}},
}
 with $d^2_N = \sum_{M \in \mathbb{Z}} d^{\alpha}_{N+M}d_{\alpha -M}$ and  $\bar{d}^2_N=\sum_{M \in \mathbb{Z}} \dbar_{\ad N+M} \dbar^{\ad}_{-M}$.
 
 We remark that when commuting or anti-commuting with functions of $\theta_0$ and $\thetab_0$, 
 \eqn{d_{0 \al} = - p_{\al 0} -  (\sigma^m \bar{\theta}_0)_{\al} \al_{0m} +..., \quad \dbar_{0 \ad} = -\bar{p}_{\ad 0} + (\theta_0 \sigma^m)_{\ad} \al_{0m}+...
}[WBDs]
act as the usual derivatives $D_{\al}$ and $\bar{D}_{\ad}$ of \cite{Wess:1992cp}. To make contact with that notation, one can use the replacements $p_{\al 0} \rightarrow - \frac{\pd}{\pd \theta^{\al}}$, $\bar{p}_{\ad 0} \rightarrow \frac{\pd}{\bar{\pd \theta}^{\ad}}$, $ \theta_0^{\al} \rightarrow \theta^{\al}$, $ \bar{\theta}_0^{\ad} \rightarrow \bar{\theta}^{\ad}$ and $\al^m_0 = -i \pd^m$. Note that $(D^{\al})^{\dagger} = \bar{D}^{\ad}$ in \cite{Wess:1992cp}, while $(d^{\al}_0)^{\dagger} = - \dbar^{\ad}_0$. This is not contradictory because $d^{\al}_0$ and $\dbar^{\ad}_0$ act through commutators/anticommutators.

\subsec{The string field $\Phi$ and superspace action}[sectionSFT]
It was shown in \cite{Berkovits:1995ab} that the string field $\Phi$ of the manifestly $SO(3,1)$ super-Poincaré invariant open superstring field theory can be written as $\Phi = \Phi_{-1} + \Phi_0 + \Phi_1$ modulo exact terms in $G^+/\tilde{G}^+$, with the field $\Phi_n$ carrying $n$ units of $\rho$-charge and $-n$ units of CY-charge. The linearized equations of motion for $\Phi$ are \begin{align}
&\tilde{G}^+_4 G^+_4 \Phi_{-1} + \tilde{G}^+_4G^+_6 \Phi_0 + \tilde{G}_4^+\tilde{G}_6^+ \Phi_1 = 0, \label{SFTeom1.1}\\
&(\tilde{G}_6^+ G_6^+ + \tilde{G}_4^+ G_4^+)\Phi_0 + \tilde{G}_4^+ G_6^+ \Phi_1 + \tilde{G}_6^+ G_4^+\Phi_{-1} = 0 \label{SFTeom},\\
&G_4^+ G_6^+ \Phi_{-1} + G_4^+ \tilde{G}_6^+ \Phi_0 + G_4^+ \tilde{G}_4^+ \Phi_1 = 0. \label{SFTeom1.3}
\end{align}
When we write $G^+_4 \mathcal{O}$ instead of  $G^+_4(z) \mathcal{O}$ we mean taking the contour integral of $G^+_4$ around $\mathcal{O}$, i.e., $ \oint dz \; G^+_4(z) \mathcal{O}$, and similarly for the other generators. 

These equations of motion are invariant under the linearized gauge transformations \begin{align}
&\delta \Phi_{-1} = G_4^+ \Lambda_{-2} + G^+_6 \Lambda_{-1} + \tilde{G}^+_6 \Lambda_0 + \tilde{G}^+_4 \Lambda_{1}, \label{SFTgauge1} \\
&\delta \Phi_0  = G_4^+ \Lambda_{-1} + G_6^+ \Lambda_0 + \tilde{G}^+_6 \Lambda_1 + \tilde{G}^+_4 \Lambda_2, \label{SFTgauge2} \\
&\delta \Phi_1  = G_4^+ \Lambda_0 + G_6^+ \Lambda_1 + \tilde{G}_6^+ \Lambda_2 + \tilde{G}_4^+ \Lambda_3. \label{SFTgauge3}
\end{align}
where the gauge parameter $\Lambda_n$ carries $n$ units of $\rho$-charge and $-n-1$ units of CY-charge. In the situation that we will encounter, $\Phi_{-1}$ and $\Phi_1$ will be algebraically gauged away. Due to this, only \eqref{SFTeom} will contribute to the quadratic superspace action \begin{equation}
S = \bra \Phi_0 (\tilde{G}_6^+ G_6^+ + \tilde{G}_4^+ G_4^+)\Phi_0 \ket,
\end{equation}
which is evaluated as a two-point CFT correlation function on the plane with the normalization $\bra e^{-i \rho + i H_C} (\theta_0 \theta_0) (\thetab_0 \thetab_0) \ket = 1$.

As an example, consider the four-dimensional massless sector of the open superstring which is independent of the compactification. Since there is nothing we can write with conformal weight zero at zero momentum for $\Phi_{1}$ and $\Phi_{-1}$, one finds that $\Phi_0 = V(x^m, \theta_0^{\al}, \bar{\theta}_0^{\ad})$ and $\Phi_1 = \Phi_{-1} = 0$ where $V$ is the  standard real vector superfield for the four-dimensional super-Maxwell multiplet. Then, schematically, the quadratic superspace action and linearized gauge transformations are \eqn{
S= \bra V \tilde{G}_4^+ G_4^+  V \ket , \quad \delta V = G_4^+ \Lambda_{-1} + \tilde{G}^+_4 \Lambda_2.
}

\newsec{Superstrings in a Constant Electromagnetic Background Field}[chargedstrings]

 This section describes the quantization of charged open superstrings in a constant electromagnetic background. Besides the usual coupling to the Lorentz current, a new boundary term $S_{b}$ is added to the worldsheet action, coupling the spacetime fermionic worldsheet variables to the background $U(1)$ gauge field in a non-minimal fashion. Expressions for the oscillator modes and also for the small $\mathcal{N}=4$ generators which generalize the free case are obtained. Charged open bosonic strings were studied in \cite{Abouelsaood:1986gd}, \cite{Argyres:1989qr}, \cite{Argyres:1989cu} and, more recently, in \cite{Porrati:2010hm}.
 
 \subsec{Worldsheet action and boundary conditions}

To couple the superstring to a constant background gauge field, we employ the hybrid formalism in terms of oscillator modes for the four-dimensional variables. However, the chiral boson $\rho$ will continue to be described using the free field OPE's in our treatment. This will preserve the four-dimensional spacetime supersymmetry and the gauge symmetry of our superstring field theory description. More importantly, our treatment will also preserve the form of the small $\mathcal{N}=4$ algebra, so that the reasoning used in constructing the formalism in the free case will also hold for the interacting case.

We will consider an open string with total charge $Q=q_0 + q_\pi$ and, as usual, the constant electromagnetic field strength $F_{mn}$ couples to the charges $q_0$ and $q_\pi$ at the ends of the string by the conserved current associated with Lorentz transformations\foot{ For a generic rank-2 tensor $T$, we use the conventions $T_{[mn]} = \frac{1}{2} ( T_{mn} - T_{nm})$ and $T_{(mn)} = \frac{1}{2} ( T_{mn} + T_{nm})$.} \eqn{
 J^{mn}_{\tau} = - \frac{i}{\al'} \pd_{\tau} x^{[m}x^{n]} - \frac{1}{2} \Big[ (p \sigma^{mn} \theta) + (\bar{p} \sigmabar^{mn} \bar{\theta}) - ( \hat{p} \sigma^{mn} \hat{\theta}) - ( \hat{\bar{p}} \sigmabar^{mn} \hat{\bar{\theta}}) \Big],
}[Lcurrent] 
obtained by varying the worldsheet action with  \eqna{
&\delta x^m = \omega^{mn} x_n, \quad \delta \theta_{\al} = -\frac{1}{2}\omega_{mn} \sigma^{mn \;\; \bt}_{ \;\;\;\;\; \al} \theta_{\bt}, \quad \delta \bar{\theta}_{\ad} =  \frac{1}{2} \omega_{mn} \sigmabar^{mn \bd}_{\;\;\;\;\;\; \ad} \bar{\theta}_{\bd}, \\
&\delta p_{\al} =- \frac{1}{2}\omega_{mn} \sigma^{mn \;\; \bt}_{ \;\;\;\;\; \al} p_{\bt}, \quad \delta \bar{p}_{\ad} =  \frac{1}{2} \omega_{mn} \sigmabar^{mn \bd}_{\;\;\;\;\;\; \ad} \bar{p}_{\bd},
}[lorentztransf]
 as well as for the right-moving fields. The variables $\omega_{mn}$ represent the parameters responsible for  Lorentz transformations and the matrices $\sigma^{mn}$, $\sigmabar^{mn}$ generate the spinor representations of the Lorentz group and are defined as in the appendix of \cite{Wess:1992cp}.
 
 The interaction term in the Euclidean action is then \eqn{
S_{int} = \frac{\al'}{2} \int d \tau F_{mn} \bigg\{ - \frac{i}{\al'} \pd_\tau x^m x^n
- \frac{1}{2} \Big[ (p\sigma^{mn} \theta) + (\bar{p} \sigmabar^{mn} \thetab) - (\hat{p} \sigma^{mn} \hat{\theta}) - (\hat{\bar{p}} \sigmabar^{mn} \hat{\bar{\theta}}) \Big]\bigg\},
}[s_int]
and the total worldsheet action is given by \eqn{
S = S_0 + q_0 S_{int} \big|_{\sigma = 0} + q_\pi S_{int} \big|_{\sigma= \pi}.
}[s_0s_int]

\subsubsec{Bosonic worldsheet fields}

We start by analyzing the known case of the bosonic variables. The equation of motion and boundary conditions for the bosonic worldsheet fields are \eqna{
& \partial_{\tau}^2 x^m + \partial_\sigma^2 x^m = 0, \\
&\pd_{\sigma} x^n + 2 \pi i \al'q_0 \pd_{\tau}x_m F^{mn} = 0, \quad \sigma = 0, \\
&\pd_{\sigma} x^n - 2 \pi i \al'q_\pi \pd_{\tau}x_m F^{mn} = 0, \quad \sigma = \pi.
}[]

This set of equations has a general solution \cite{Argyres:1989qr} \eqn{
x(w) = \frac{\widehat{x}}{2} - \bigg( \frac{1}{QF} \bigg) \widehat{p} + \sqrt{\frac{\al'}{2}} \sum_N \bigg( \frac{1}{-\varepsilon -iN} \bigg) \bigg[ e^{(-\varepsilon -iN)(-i \tau + \sigma) + \gamma} + e^{(-\varepsilon -iN)(-i \tau - \sigma) - \gamma} \bigg] \al_N, \label{x_msolution}
}[]
where we are using a matrix notation for the spacetime indices, for example, $(F^{-1}) \widehat{p} = (F^{-1})^{mn} \widehat{p}_n$, and the real antisymmetric spacetime matrices $\varepsilon$, $\gamma$ and $\gamma'$ are defined as \eqna{
&\varepsilon = \frac{1}{\pi} (\gamma + \gamma'), \\
&\gamma = \text{arctanh}(2 \pi \al' q_0 F), \\
&\gamma'= \text{arctanh}(2\pi \al'q_\pi F).
}[]

Canonical quantization implies for the coefficients $\widehat{x}^{\;m}$, $\widehat{p}^{\;m}$ and $\al_N^m$ the following commutators \eqn{
[ \al_M^m, \al_N^n] = (M - i \varepsilon)^{mn} \delta_{M+N,0}, \quad [\widehat{x}^{\;m}, \widehat{p}^{\;n}] = i \eta^{mn}.
}[]

As the notation suggests, $\widehat{x}^{\;m}$ and $\widehat{p}^{\;m}$ have the interpretation of physical position and momentum, as is easily verified analyzing the point particle limit. This interpretation also justifies the factor of $\frac{1}{2}$ in $\widehat{x}^{\;m}$, which is explained in \cite{Argyres:1989qr}. In particular, one could expect that $\al_0^m$ would be the covariant derivative for the charged string, but note that $ [\al^m_0, \al^n_0] = - i \varepsilon^{mn}$, which is not the commutation relation of a covariant derivative, namely, $[\mathcal{D}_m , \mathcal{D}_n]= iQ F_{mn}$. However, suppose we construct a real spacetime matrix $\mathcal{M}$ satisfying \eqn{
\mathcal{M}  \mathcal{M}^T = \frac{\varepsilon}{QF}.
}[Mrelation]One can then define
\eqn{
\al_0 = \mathcal{M}  (p -\frac{1}{2}QFx) = -i \mathcal{M} \mathcal{D} \equiv -i \mathfrak{D},
}[covDbosonic]
where $\mathcal{D}_n = (\pd - \frac{i}{2} QF x)_n$ is the covariant derivative with $A_m = - \frac{1}{2} F_{mn}x^n$. We then obtain \eqn{
[\al_{0m} ,\al_{0r}] =- \mathcal{M}_{mn} \mathcal{M}_{rs} [\mathcal{D}^n,\mathcal{ D}^s] = -i Q (\mathcal{M}  F  \mathcal{M}^T)_{mr} = -iQ\left(\frac{\varepsilon}{QF}F\right)_{mr} = -i \varepsilon_{mr},
}[]
as desired.

In the neutral string limit, $Q \rightarrow 0$, \Mrelation\ implies that \eqn{
\mathcal{M}  \mathcal{M}^T \rightarrow (1 - (\pi q_0 F)^2)^{-1}. \label{M1}
}[]
Consequently, it is consistent for small $F_{mn}$ to define \eqn{
\mathcal{M} \rightarrow  (1 + \pi q_0 F)^{-1}.
}[M2]

Moreover, we will assume that $\varepsilon$, $\mathcal{M}$ and $\gamma$ commute as matrices, which is justified given that each can be expressed in a power series in $F$ and, as such, can be put in block skew-diagonal form by a suitable Lorentz transformation.

\subsubsec{Fermionic worldsheet fields}

Now we turn to the fermionic worldsheet fields. As in the bosonic case, the equations of motion are unaffected by boundary contributions. The subtle part is to find solutions that satisfy the non-trivial boundary conditions.  We will now argue that the boundary conditions implied by the interaction term of \eqref{s_0s_int} are inconsistent, and that preservation of the worldsheet superconformal invariance will require additional interaction terms for the fermionic worldsheet fields.

In the strip, at $\sigma = \pi$, the boundary condition from \eqref{s_0s_int} that the fermionic fields would need to satisfy is \eqn{
p^{\al} \delta \theta_{\al} + \bar{p}_{\ad} \delta \thetab^{\ad} + \hat{p}^{\al} \delta \hat{\theta}_{\al} + \hat{\bar{p}}_{\ad} \delta \hat{\bar{\theta}}^{\ad} - \frac{\pi \al' q_{\pi}}{2}  \bigg[  \delta (p F \cdot \sigma \theta) + \delta  ( \bar{p} F \cdot \sigmabar \thetab) - \delta  ( \hat{p} F \cdot \sigma \hat{\theta}) - \delta (\hat{\bar{p}} F \cdot \sigmabar \hat{\bar{\theta}})\bigg] =0,
}[fermibc]
where $F \cdot \sigma \equiv F_{mn} \sigma^{mn}$, and the boundary condition at $\sigma=0$ would have a similar form but with the replacement $q_\pi \rightarrow -q_0$. One would
then need to find an expression for the worldsheet fields that satisfies \fermibc, but it turns out that a solution of the form (suppressing anti-chiral fields) \eqn{
\hat{p}^{\al} = a p^{\al} + b F_{mn} (p \sigma^{mn})^{\al}, \quad \hat{\theta}^{\al} = c \theta^{\al} + d F_{mn} (\theta \sigma^{mn})^{\al}
}
cannot be found for any non-trivial value of the coefficients $(a,b,c, d)$. This suggests that we need to modify the boundary term in the action for the fermionic fields. 

To solve this problem, we will ignore the worldsheet action for a moment and start with a requirement that fixes the boundary conditions of the fermionic fields.  It will then be realized that a new boundary contribution $S_b$, in addition to the terms in equation \eqref{s_int}, needs to be added to the action for these boundary conditions to be satisfied.

Note that in \cite{Berkovits:2002ag} the equations of motion for the supersymmetric Born-Infeld theory were obtained by demanding that $\hat{\mathcal{Q}} = \mathcal{Q}$ at the boundary, where $\mathcal{Q}$ and $\hat{\mathcal{Q}}$ are the left and right-moving BRST operators. Following the same logic, we want to impose boundary conditions for the fermionic worldsheet fields such that the left and right-moving small $\mathcal{N}=4$ superconformal generators coincide at $\sigma=0$ and $\sigma= \pi$. This will be accomplished by first looking at the terms in the superconformal generators that have $ \pd x^m$, which is already fixed by \eqref{x_msolution}. To be more concrete, from \eqref{x_msolution}, we have (recall that $z= e^{-iw}$, with $w = \sigma + i \tau$)\eqna{
&\pd_w x^m (w) = - \sqrt{\frac{\al'}{2}} \sum_N \Big[ e^{( -\varepsilon -i N)(-\sigma -i \tau) - \gamma} \Big]^{m}_{\;\;n} \al_N^n, \\
&\pd_{\bar{w}} x^m( \bar{w}) = \sqrt{\frac{\al'}{2}} \sum_N \Big[ e^{(-\varepsilon -iN)(\sigma -i \tau) + \gamma} \Big]^m_{\;\;n} \al_N^n.
}[xbackmodes]
Therefore, at the boundary, we obtain the following relations
\eqn{ \pd_{\bar{w}}x^m(\bar{w})\biggr|_{\sigma=0} = -[e^{2 \gamma}]^m_{\;\;n} \pd_w x^n(w)\biggr|_{\sigma=0}, \quad \pd_{\bar{w}}x^m(\bar{w})\biggr|_{\sigma=\pi} = -[e^{-2 \gamma'}]^m_{\;\;n} \pd_w x^n(w)\biggr|_{\sigma=\pi}.  }[pdxbc]

Notice that the boundary conditions relating $\bar{\pd}x^m$ to $\pd x^m$ resemble a Lorentz transformation given by choosing $\omega = 2 \gamma$ at $\sigma = 0$ and $\omega = - 2 \gamma'$ at $\sigma=\pi$ in view of the exponentiated form of equations \eqref{lorentztransf}. Besides that, we know that the small $\mathcal{N}=4$ superconformal generators are composed of Lorentz invariant terms, with the left-moving ones containing the worldsheet fields $(\pd x^m, p_{\al}, \theta_{\al}, \bar{p}_{\ad}, \bar{\theta}_{\ad})$ and the right-moving ones containing $(\bar{\pd} x^m, \hat{p}_{\al}, \hat{\theta}_{\al}, \hat{\bar{p}}_{\ad}, \hat{\bar{\theta}}_{\ad})$. So to obtain the intended boundary conditions relating the left and right-moving superconformal generators in the interacting case, namely,
\eqn{\hat{G}^{\pm}(\bar{z})=G^{\pm}(z), \quad \hat{\tilde{G}}^{\pm}(\bar{z})=\tilde{G}^{\pm}(z), \quad \hat{T}(\bar{z})= T(z),
}[]
at $\text{Im}\{z\} = 0$, one can relate the left and right-moving fermionic worldsheet fields by a condition resembling a Lorentz transformation given with the same parameter $\omega_{mn}$ as in \eqref{pdxbc}, i.e., at $\sigma = 0$ \begin{align}
&\hat{p}_{\al}(\bar{w}) =-  [e^{- \gamma \cdot \sigma}]_{\al}^{\; \bt} p_{\bt}(w), \quad \hat{\bar{p}}_{\ad}(\bar{w}) =- [e^{ \gamma \cdot \sigmabar}]^{\bd}_{\; \ad} \bar{p}_{\bd}(w), \label{bcfermi1}\\
& \hat{\theta}_{\al}(\bar{w}) = [e^{ - \gamma \cdot \sigma}]_{\al}^{\; \bt} \theta_{\bt}(w), \quad \hat{\bar{\theta}}_{\ad}(\bar{w}) = [e^{ \gamma \cdot \sigmabar}]^{\bd}_{\; \ad} \bar{\theta}_{\bd}(w). \label{bcfermi2}
\end{align}
And at $\sigma = \pi$ \begin{align}
&\hat{p}_{\al}(\bar{w}) =  -[ e^{\gamma' \cdot \sigma}]_{\al}^{\; \bt} p_{\bt} (w), \quad \hat{\bar{p}}_{\ad}(\bar{w}) = - [e^{ - \gamma' \cdot \sigmabar}]^{\bd}_{\; \ad} \bar{p}_{\bd}(w), \label{bcfermi3} \\
& \hat{\theta}_{\al}(\bar{w}) = [e^{ \gamma' \cdot \sigma}]_{\al}^{\; \bt} \theta_{\bt}(w), \quad \hat{\bar{\theta}}_{\ad}(\bar{w})= [e^{- \gamma'\cdot \sigmabar}]^{\bd}_{\; \ad} \bar{\theta}_{\bd}(w), \label{bcfermi4}
\end{align}
where $(\gamma \cdot \sigma)_{\bt}^{\; \al} \equiv \gamma_{mn} \sigma^{mn \;\; \al}_{\;\;\;\;\; \bt}$, $(\gamma \cdot \sigmabar)^{\ad}_{\; \bd} \equiv \gamma_{mn} \sigmabar^{mn \ad}_{\;\;\;\;\;\; \bd}$ and similarly for $\gamma'$. We stress that whenever the letter $\sigma$ ($\sigmabar$) is accompanied by a \textit{dot}, we mean the Lorentz generator $\sigma^{mn}$ ($\sigmabar^{mn}$) and not the worldsheet variable $\sigma$. We hope the different index structure helps to avoid any confusion. 

By the same reasoning, the mode expansions of the fermionic fields should take the following form \eqna{
& p_{\al}(w) = - i\sum_{N} \Big[ e^{(-iN +\frac{1}{2}\varepsilon\cdot\sigma) (-\sigma -i \tau) + \frac{1}{2}\gamma\cdot\sigma }  \Big]_{\al}^{\;\;\bt}p_{\bt N},
\\
&\theta_{\al}(w) = \sum_{N}\Big[ e^{(-iN +\frac{1}{2}\varepsilon\cdot\sigma)(-\sigma -i \tau) + \frac{1}{2}\gamma\cdot\sigma }  \Big]_{\al}^{\;\;\bt}\theta_{\bt N},
\\
& \bar{p}_{\ad}(w) = - i\sum_{N} \Big[ e^{(-iN - \frac{1}{2}\varepsilon\cdot\sigmabar)(-\sigma -i \tau) - \frac{1}{2}\gamma\cdot\sigmabar }  \Big]_{\;\;\;\ad}^{\bd} \bar{p}_{\bd N},
\\ 
&\bar{\theta}_{\ad}(w) = \sum_{N} \Big[ e^{(-iN - \frac{1}{2} \varepsilon\cdot \sigmabar)(-\sigma -i \tau) - \frac{1}{2} \gamma\cdot \sigmabar }  \Big]_{\;\;\;\ad}^{\bd} \bar{\theta}_{\bd N},
}[]
so that the exponential factors of the background are not present in the Lorentz invariant terms of the small $\mathcal{N}=4$ superconformal generators. Note that $(\varepsilon \cdot \sigma)_{\bt}^{\; \al} \equiv \varepsilon_{mn} \sigma^{mn \;\; \al}_{\;\;\;\;\; \bt}$ and $(\varepsilon \cdot \sigmabar)^{\ad}_{\; \bd} \equiv \varepsilon_{mn} \sigmabar^{mn \ad}_{\;\;\;\;\;\; \bd}$.

In short, one can say that Lorentz invariance fixed the form of the fermionic worldsheet fields in the interacting case. The appearance of a Lorentz transformation in the boundary conditions linking the left and right-moving fields is not surprising from the point of view of an open string with endpoints of charge $q_0$ and $q_\pi$ attached to a D-brane. In this setting, T-duality can be used to show that the tilt of the D-brane in spacetime is related to the field strength $F$, consequently, the boundary conditions express the fact that the coordinates become rotated by the gauge field in the dual description.

Using equations \eqref{bcfermi1} to \eqref{bcfermi4} in \eqref{fermibc}, one can show that the following term should be added to the action for it to imply our desired boundary conditions \eqna{
S_b &= q_0 \frac{\al '}{8} \int d \tau F_{mn} \Big[-(\hat{p} \{e^{-\gamma \cdot \sigma}, \sigma^{mn} \} \theta) + (p \{e^{\gamma \cdot \sigma}, \sigma^{mn} \} \hat{\theta}) \\
& \quad \quad \quad \quad \quad \quad \quad \quad - (\hat{\bar{p}} \{e^{-\gamma \cdot \sigmabar}, \sigmabar^{mn}\} \bar{\theta}) + (\bar{p} \{e^{\gamma \cdot \sigmabar}, \sigmabar^{mn}\} \hat{\thetab}) \Big]\biggr|_{\sigma=0} \\
&+ q_\pi \frac{\al'}{8} \int d\tau F_{mn} \Big[ -(\hat{p}\{ e^{\gamma'\cdot \sigma}, \sigma^{mn}\} \theta) + (p \{ e^{-\gamma' \cdot \sigma}, \sigma^{mn} \} \hat{\theta}) \\
&  \quad \quad \quad \quad \quad \quad \quad \quad - (\hat{\bar{p}} \{e^{\gamma' \cdot \sigmabar}, \sigmabar^{mn} \} \bar{\theta}) + (\bar{p} \{e^{ -\gamma '\cdot \sigmabar}, \sigmabar^{mn} \} \hat{\thetab}) \Big]\biggr|_{\sigma=\pi}. 
}[s_b]
Up to the addition of trivial terms whose interaction term vanishes, the contribution $S_b$ is the unique symmetric combination in $(p_{\al}, \theta_{\al}, \bar{p}_{\ad}, \bar{\theta}_{\ad})$ and $(\hat{p}_{\al}, \hat{\theta}_{\al}, \hat{\bar{p}}_{\ad}, \hat{\bar{\theta}}_{\ad})$ that we can add to the action to obtain the boundary conditions of  \eqref{bcfermi1} to \eqref{bcfermi4}.

So gathering expressions \eqref{s_0}, \eqref{s_int} and \eqref{s_b}, we can write the total worldsheet action for the charged open superstring in the hybrid formalism coupled to a constant electromagnetic background as

\eqn{
S = S_0 + q_0 S_{int} \big|_{\sigma = 0} + q_\pi S_{int} \big|_{\sigma= \pi} + S_b.
}[totalaction_int]

\subsec{Superconformal generators and commutation relations}

We will present here the description of the superconformal generators in terms of oscillator modes, which is central for the computations contained in future sections.

Canonical quantization of the fermionic worldsheet variables implies the same commutation relations as in the free case for the fermionic modes \eqn{
\{ p^{\bt}_M, \theta_{\al N}\} =\delta^{\bt}_{\al} \delta_{M + N, 0}, \quad \{ \bar{p}_{\bd M}, \bar{\theta}^{\ad}_N \} = \delta^{\ad}_{\bd}  \delta_{M + N, 0}.
}[]
The boundary conditions and mode expansions determined that the Lorentz invariant terms present in the superconformal generators, such as $d^2(z)$, are holomorphic and independent of exponential factors of the background gauge field. This makes it straightforward to obtain the modes of $(G^{\pm}, \tilde{G}^{\pm},T)$ through the usual method of contour integration, applying the doubling trick to consider only the left-moving variables defined in the whole plane.

One starts with the mode expansions in the plane \eqna{
&\pd x^m = -i \sqrt{\frac{\al'}{2}} \sum_N [z^{-i \varepsilon -N -1} e^{-\gamma}]^m_{\;\;n} \al_N^n, \\
&p_{\al}(z) = \sum_N [z^{-N - \frac{i}{2} \varepsilon \cdot \sigma} e^{\frac{1}{2} \gamma \cdot  \sigma}]_{\al}^{\; \bt} p_{\bt N}, \\
&\theta_{\al}(z)= \sum_N [z^{-N - \frac{i}{2} \varepsilon \cdot \sigma} e^{\frac{1}{2} \gamma \cdot  \sigma}]_{\al}^{\; \bt} \theta_{\bt N}, \\
&\bar{p}_{\ad}(z) = \sum_N [z^{-N + \frac{i}{2} \varepsilon \cdot \sigmabar} e^{- \frac{1}{2}\gamma \cdot \sigmabar}]^{\bd}_{\;\ad} \bar{p}_{\bd N}, \\
&\bar{\theta}_{\ad}(z) = \sum_N [z^{-N + \frac{i}{2} \varepsilon \cdot \sigmabar} e^{- \frac{1}{2}\gamma \cdot \sigmabar}]^{\bd}_{\;\ad} \bar{\theta}_{\bd N}. 
}[modeplaneback]
Then, as an example, substituting equations \eqref{modeplaneback} in $d^2(z)$ and $\dbar^2(z)$, where $d_{\al}$ and $\dbar_{\ad}$ were given in \eqref{susyvariables}, we obtain
\eqn{
d^2(z) = \sum_N \frac{d^2_N}{z^{N+2}}, \quad \dbar^2(z)= \sum_N \frac{\dbar^2_N}{z^{N+2}},
}[]
which have the same form as in the free case, with $d^2_N = \sum_{M \in \mathbb{Z}} d^{\alpha}_{N+M}d_{\alpha -M}$ and  $\bar{d}^2_N=\sum_{M \in \mathbb{Z}} \dbar_{\ad N+M} \dbar^{\ad}_{-M}$. And the modes of the supersymmetric variables can be read off by substituting the expressions \modeplaneback in \susyvariables \eqna{
&d_{\al N} = -p_{\al N} - \sum_R (\sigma^m \thetab_R)_{\al} \al_{m N-R} - \sum_{RS}(N-2R-S) (\thetab_R \thetab_S) \theta_{\al N-R-S} \\
& \quad \; \; \, - \frac{i}{2}  \sum_{RS} (\thetab_R \thetab_S) (\varepsilon \cdot \sigma)_{\al}^{\;\bt} \theta_{\bt N-R-S},\\
&\bar{d}_{\ad N} = - \bar{p}_{\ad N} + \sum_R (\theta_R \sigma^m)_{\ad} \al_{m N-R} - \sum_{RS} (N-2R -S) (\theta_R \theta_S) \bar{\theta}_{\ad N-R-S} \\
& \quad \; \; \, + \frac{i}{2}  \sum_{RS} (\theta_R \theta_S) (\varepsilon \cdot \sigmabar)^{\bd}_{\; \ad} \thetab_{\bd N-R-S}, \\
& \Pi_{\al \ad N} = -i \sigma^m_{\al \ad} \al_{m N} + 2i \sum_R (N -2R) \theta_{\al R} \thetab_{\ad N-R} + \sum_R (\varepsilon \cdot \sigma)_{\al}^{\;\bt} \theta_{\bt R} \thetab_{\ad N-R} \\
& \quad \quad  \; \, +  \sum_R (\varepsilon \cdot \sigmabar)^{\bd}_{\;\ad} \theta_{\al R} \thetab_{\bd N-R},\\
& \pd \theta_{\al N} = -N \theta_{\al N} -\frac{i}{2}  (\varepsilon \cdot \sigma)_{\al}^{\;\bt} \theta_{\bt N}, \quad \pd \thetab_{\ad N} = -N \thetab_{\ad N} + \frac{i}{2}  (\varepsilon \cdot \sigmabar)^{\bd}_{\;\ad} \thetab_{\bd N} .
}[modevariables]

Note that $\pd \theta_{\al 0} = \pd \thetab_{\ad 0} = 0$ when $\varepsilon = 0$. The supersymmetric modes of the interacting fields satisfy \eqna{
&[d_{\al N} ,\Pi_{\bt \bd M}] = 4i \epsilon_{\al \bt} \pd \bar{\theta}_{\bd N+M}, \quad [\bar{d}_{\ad N}, \Pi_{\bt \bd M}]= -4i \epsilon_{\ad \bd} \pd \theta_{\bt N+M}, \\
& \{d_{\al M}, \pd \theta^{\bt}_N\}= -N \delta^{\bt}_{\al} \delta_{M+N,0} + \frac{i}{2}  (\varepsilon \cdot \sigma)_{\al}^{\; \bt} \delta_{M+N,0}, \\
& \{\bar{d}^{\ad}_M, \pd \bar{\theta}_{\bd N}\}= -N \delta^{\ad}_{\bd} \delta_{M+N,0} + \frac{i}{2}  (\varepsilon \cdot \sigmabar)^{\ad}_{\: \bd} \delta_{M+N,0}, \\
& \{d_{\al N},\bar{d}_{\ad M}\}= 2i \Pi_{\al \ad M+N}.
}[commback]
These commutation relations follow from equation \modevariables by using
\eqn{
 [\al^m_M, \al^n_N] = (M - i \varepsilon)^{mn} \delta_{M+N,0}, \quad \{ p^{\bt}_M, \theta_{\al N}\} =\delta^{\bt}_{\al} \delta_{M + N, 0}, \quad \{ \bar{p}_{\bd M}, \bar{\theta}^{\ad}_N \} = \delta^{\ad}_{\bd}  \delta_{M + N, 0},
}[]
and preserve the same structure as the commutation relations in the free case \eqref{commfreeSusy}. Observe that, from \eqref{modevariables}, the introduction of the background modified the \textit{super derivatives} zero modes by a term proportional to $\varepsilon$ \eqna{
&d_{ \al 0}  f(\theta_0, \thetab_0)= \big[- p_{\al 0} - (\sigma^m \bar{\theta}_0)_{\al} \al_{0m} - \frac{i}{2} (\bar{\theta}_0 \bar{\theta}_0) (\varepsilon \cdot \sigma)_{\al}^{\; \bt} \theta_{\bt 0}\big] f(\theta_0, \thetab_0), \\
&\dbar_{ \ad 0} f(\theta_0, \thetab_0) = \big[-\bar{p}_{\ad 0} + (\theta_0 \sigma^m)_{\ad} \al_{0m} + \frac{i}{2} (\theta_0 \theta_0) (\varepsilon \cdot \sigmabar)^{\bd}_{\; \ad} \bar{\theta}_{\bd 0}\big] f(\theta_0, \thetab_0), \\
&\Pi^m_0 f(\theta_0, \thetab_0) = \big[- i \al_0^m + \frac{i}{2} \epsilon^{mnrs} \varepsilon_{rs} (\theta_0 \sigma_n \bar{\theta}_0)\big] f(\theta_0, \thetab_0),
}[szeromodesint]
where $\epsilon^{mnpq}$ is the four-dimensional Levi-Civita symbol with $\epsilon_{0123} = -1$. Equation \szeromodesint can be seen as a generalization of \eqref{covDbosonic} from bosonic strings to the supersymmetric counterpart in four dimensions.

Of course, the superconformal generators have the same form as in the non-interacting case \eqna{
&G^+_4(z) = \frac{1}{2\sqrt{8}} e^{i \rho}(z)  \sum_N \frac{d^2_N}{z^{N+2}}, \quad G_4^-(z) = - \frac{1}{2\sqrt{8}} e^{-i \rho}(z) \sum_N \frac{\dbar^2_N}{z^{N+2}}, \\
&\tilde{G}^+_4(z) = - \frac{1}{2\sqrt{8}} e^{-2i \rho + iH_C}(z) \sum_N \frac{\dbar^2_N}{z^{N+2}}, \quad \tilde{G}^-_4(z) = - \frac{1}{2\sqrt{8}} e^{2i \rho -i H_C}(z)  \sum_N \frac{d^2_N}{z^{N+2}}, \\
&T_4(z) = \sum_N \frac{L_N}{ z^{N + 2}} + \frac{1}{2} \pd \rho \pd \rho(z) - \frac{i}{2} \pd^2 \rho(z), \\
&L_M = \sum_N \bigg( \frac{1}{4} \Pi^{\ad \al}_N \Pi_{\al \ad -N + M} + \pd \theta^{\al}_N d_{\al -N + M} + \pd \bar{\theta}_{\ad N} \dbar^{\ad}_{-N + M} \bigg),
}[back_gen]
 using \modevariables for the supersymmetric modes. Also, using the fact that $L_1 L_{-1} |0\ket - L_{-1} L_1 |0\ket = 2L_0 | 0 \ket $, one finds that $L_0$ does not acquire a normal ordering constant when the background field is nonzero. In the bosonic string case \cite{Argyres:1989qr}, a nonvanishing normal ordering constant was found to be proportional to $\varepsilon^2$. The vanishing of the normal ordering constant in our analysis is a consequence that our description preserves spacetime supersymmetry.

For later use, we also define the constant matrices \eqn{
\Delta_{\al}^{\; \bt} \equiv \{d_{\al 1}, \pd \theta^{\bt}_{-1}\} = \delta^{\bt}_{\al} + \frac{i}{2}  (\varepsilon \cdot \sigma)^{\;\;\bt}_{\al}, \quad \bar{\Delta}^{\ad}_{\;\bd} \equiv \{\dbar^{\ad}_{1},\pd \bar{\theta}_{\bd -1} \} = \delta^{\ad}_{\bd} + \frac{i}{2} (\varepsilon \cdot \sigmabar)^{\ad}_{\;\;\bd}.
}[]
Identities which will be useful for future computations of commutation relations between the modes can be found in appendix \ref{spinorconventions} together with our conventions for sigma matrices and spinorial indices.

\newsec{Massless Spin-1 Multiplet in a Constant Electromagnetic Background} \label{spin1section}
 
Using the description of the hybrid formalism coupled to a constant U(1) background gauge field developed in the last section, we calculate in this section the action for the four-dimensional super-Maxwell multiplet of the charged open superstring. 

\subsec{Equations of motion and superspace action}
 
 As we saw in section \ref{sectionSFT}, the vertex operator for the compactification-independent massless states of the open superstring has vanishing $J$ charge and weight zero at zero momentum and is described by a superfield $V$. In the case of charged strings, we need to allow the vector superfield $V$ describing the super-Maxell multiplet to be complex. As usual, $V$ can be expanded in terms of $\theta_0^{\al}$ and $\bar{\theta}^{\ad}_0$
 \eqna{
V(x^m, \theta_0^{\al}, \bar{\theta}_0^{\ad})& = \phi + i (\theta_0 \chi_1) - i ( \bar{\theta}_0\bar{\chi}_2) + i (\theta_0 \theta_0) M_1 - i (\bar{\theta}_0 \bar{\theta}_0) M^*_2 - (\theta_0 \sigma^m \bar{\theta}_0) A_m  \\
&- i (\bar{\theta}_0 \bar{\theta}_0) ( \theta_0 \psi_1)+ i (\theta_0 \theta_0) (\bar{\theta}_0 \bar{\psi}_2) + \frac{1}{2} (\theta_0 \theta_0) (\bar{\theta}_0 \bar{\theta}_0) D,
}[]
where $(\phi,A_m, D)$ are complex.
The equation of motion and gauge transformations for $V$ are
 \eqn{
 G^+_4 \tilde{G}^+_4 V = 0, \quad \delta V = G_4^+ \Lambda_{-1} + \tilde{G}^+_4 \Lambda_2,
}[]
 where $\Lambda_{-1}$ and $\Lambda_2$ can be written as $\Lambda_{-1} =\sqrt{8} e^{-i \rho} i \xi$ and $\Lambda_2= \sqrt{8}e^{2 i \rho - iH_c} i \zeta$, with $\xi$ and $\zeta$ carrying no $J$ charge and having conformal weight zero, i.e., they are complex vector superfields and functions of $x^m$, $\theta_0^{\al}$ and $\bar{\theta}_0^{\ad}$.
 
From \eqref{back_gen}, the equation of motion and gauge transformations in terms of the supersymmetric modes read
\eqn{
\frac{1}{2} d_1^2 \dbar_{-1}^2 V - d_0^2 \dbar_0^2 V = 0 , \quad \delta V = \frac{i}{2}d_0^2 \xi - \frac{i}{2} \dbar_0^2 \zeta.
}[]
 We can express it in terms of zero modes using that $d_1^2 \dbar_{-1}^2 V = (2 d_0^2 \dbar_0^2  -2 d^{\al}_0 \dbar_0^2 d_{\al 0}  -32 d^{\al}_0 \pd \theta_{\al 0}) V$. The equation of motion then simplifies to
 \eqn{
 d^{\al}_0 \dbar_0^2 d_{\al 0} V + 16 \pd \theta^{\al}_0 d_{\al 0} V = 0. \label{EOMmaxwell}
}[]
 The only non-trivial gauge transformation comes from $\zeta$, and one sees that \eqref{EOMmaxwell} is gauge invariant by noting that $d^{\al}_0 \dbar_0^2 d_{\al 0} \delta V = 8i d^{\al}_0 \pd \theta_{\al 0} \dbar_0^2 \zeta$. 
 
 In the free case, $\varepsilon = 0$, $d^{\al}_0$ reduces to the usual super derivatives \eqref{WBDs} and $\pd \theta^{\al}_0 =0$. So we recover the super-Maxell equation of motion $d^{\al}_0 \dbar_0^2 d_{\al 0} V=0$, or $D^{\al} \bar{D}^2 D_{\al} V =0$ in the notation of \cite{Wess:1992cp}. 
 
 Equation \eqref{EOMmaxwell} comes from the action given by evaluating $\bra V \tilde{G}_4^+ G_4^+  V \ket $, which we write in $\mathcal{N}=1$ superspace as
 \eqn{
 S=\ \frac{1}{16} \int d^4x \; p_0^2 \bar{p}_0^2 \Big[ V^\dagger (d^{\al}_0 \dbar_0^2 d_{\al 0} + 16 \pd \theta^{\al}_0 d_{\al 0})V \Big].
}[]
To get the expression in terms of components, one can use the gauge transformations to go to the WZ gauge in which the only nonzero components of $V$ and $V^\dagger$ are
\eqna{
&V = - (\theta_0 \sigma^m \bar{\theta}_0) A_m  - i (\bar{\theta}_0 \bar{\theta}_0) (\theta_0 \psi_1) + i (\theta_0 \theta_0) (\bar{\theta}_0 \bar{\psi}_2) + \frac{1}{2} (\theta_0 \theta_0) (\bar{\theta}_0 \bar{\theta}_0) D, \\
& V^\dagger= -(\theta_0 \sigma^m \bar{\theta}_0) A_m^* + i (\theta_0 \theta_0) (\bar{\theta}_0 \bar{\psi}_1) - i (\bar{\theta}_0 \bar{\theta}_0) (\theta_0 \psi_2) + \frac{1}{2} (\theta_0 \theta_0) (\bar{\theta}_0 \bar{\theta}_0)D^*,
}[]
where we also wrote $V^\dagger$ to emphasize that $V$ is a complex superfield for the charged superstring. 

Expanding the oscillator modes in \eqref{EOMmaxwell} using \eqref{modevariables}, we obtain the equations of motion for the components \begin{align}
&D=0, \\
&\mathfrak{D}_m (\sigmabar^m \psi_1)^{\ad} = 0, \\
& \mathfrak{D}_m (\sigma^m \bar{\psi}_2)_{\al} = 0, \\
&\mathfrak{D}^2 A^m - \mathfrak{D}^m \mathfrak{D}^n A_n + 2i \varepsilon^{mn}A_n = 0, \label{EOMAm}
\end{align}
with gauge transformations
\eqna{
& \delta A_m = \mathfrak{D}_m a, \\
&\delta D = \delta \psi_{1\al} = \delta \bar{\psi}^{\ad}_2 = 0,
}[]
where $a$ is an arbitrary gauge parameter and recall that $ [\mathfrak{D}_m, \mathfrak{D}_n] = - [\al_{0m}, \al_{0n}] = i \varepsilon_{mn}$. These equations of motion are obtained by varying the action
\eqn{
S = \frac{1}{2} \int d^4 x \Big[ A^*_m \left( \mathfrak{D}^2 A^m - \mathfrak{D}^m \mathfrak{D}^n A_n + 2i \varepsilon^{mn}A_n \right) -i (\bar{\psi}_1 \sigmabar^m \mathfrak{D}_{m} \psi_1) - i(\psi_2 \sigma^m \mathfrak{D}_{m} \bar{\psi}_2) + D^* D \Big].
}[]
Note that when the background is zero  ($\varepsilon = 0$), the above action becomes two decoupled actions for the super-Maxwell multiplet 
\eqn{
S= \int d^4 x \Big[ - \frac{1}{4} F^2_1  - \frac{i}{2} (\bar{\psi}_1 \sigmabar^m \pd_m \psi_1) + \frac{1}{2} D^2_1 + ( 1 \leftrightarrow 2) \Big],
}[]
where $A= A_1 + i A_2$, $D= D_1 + i D_2$ and $F_{Imn} = \pd_m A_{In} - \pd_n A_{Im}$ ($I=1,2$).

If we perform the substitution $A_m \rightarrow (\mathcal{M} A)_m$, \eqref{EOMAm} and the gauge transformation for $A_m$ can be put in the form given by \cite{Argyres:1989qr}
\eqn{
( \mathcal{D}\cdot \frac{\varepsilon}{QF} \cdot\mathcal{D} )A^{m} - \mathcal{D}^{m}(\mathcal{D}\cdot \frac{\varepsilon}{QF} \cdot A) 
+ 2i(\varepsilon A)^{m} =0, \quad \delta A_m = \mathcal{D}_m a. \label{spin1Nappi}
}[]

Observe that the vector field $A_m$ remains massless in the presence of the background. This is due to the normal ordering constant being absent in the Virasoro algebra of the four-dimensional part of the superstring \eqref{back_gen}, an effect of our supersymmetric description. This can be contrasted with the results found in \cite{Argyres:1989qr}, where $A_m$ acquires a mass term proportional to $\varepsilon^2$, a consequence of the shift in the normal ordering constant by the same amount. Nevertheless, one can check that the difference between equation \eqref{spin1Nappi} and equation (3.5) of \cite{Argyres:1989qr} has vanishing gauge variation, so both results are consistent. Notice when comparing \eqref{spin1Nappi} with (3.5) of \cite{Argyres:1989qr} that there is a sign difference in the term with no derivatives because we define the commutator $[\mathcal{D}_m, \mathcal{D}_n] = i Q F_{mn}$, whereas $[\mathcal{D}_m, \mathcal{D}_n] = -i Q F_{mn}$ in \cite{Argyres:1989qr}.

When  $q_\pi \rightarrow - q_0$, or the neutral string limit, one has
\eqn{
\Big[ \pd \cdot (1 - \pi^2q^2_0F^2)^{-1} \cdot\pd \Big]A^{m}_1 - \pd^{m}\Big[ \pd \cdot  (1 - \pi^2q^2_0F^2)^{-1} \cdot A_1 \Big] = 0, \label{BIAm}
}[]
and
 \eqn{
  [(1+ \pi q_0 F)^{-1}]_{mn} \bar{\sigma}^{m}\pd^{n} \psi_1  = 0, \label{BIspinor}
}[]
where we used \eqref{M1} and \eqref{M2}. Similar expressions hold for $A_{2m}$ and $\psi_2$.
Equations \eqref{BIAm} and \eqref{BIspinor} can be obtained by varying the supersymmetric Born-Infeld action \cite{Fradkin:1985qd} \cite{Born:1934gh}
\eqn{
S_{superBI} = \int d^4 x \Big[ - \text{det} \Big( \eta_{mn} + q_0 \pi F_{1 mn} - 2 (\bar{\psi}_1 \sigma_m \pd_n \psi_1) \Big) \Big]^{1/2}. \label{BIaction}
}[]
More precisely, as shown in \cite{Argyres:1989qr}, equation \eqref{BIAm} can be obtained by expanding the field strength around a constant background in the equations of motion coming from \eqref{BIaction}. 

\newsec{Superspace Action of the First Massive State of the Superstring in a Constant Electromagnetic Background}[actionmass]

Repeating the steps of the last section, we compute here the superspace action for the first massive states of the charged open superstring compactified to four dimensions and coupled to a constant electromagnetic background. This action is non-polynomial in $F_{mn}$ and describes a massive complex spin-2 multiplet and two massive complex scalar multiplets, which are the compactification-independent states in four dimensions preserving $\mathcal{N}=1$ supersymmetry. The case without an electromagnetic background was studied in  \cite{Berkovits:1998ua}.

 \subsec{String field/vertex operator}
 
 Since we are ignoring  compactification-dependent contributions, fields that depend on the internal directions $j$ of the Calabi-Yau are not allowed and the most general complex string field having conformal weight +1 at zero momentum and $(\text{mass})^2 = \frac{1}{\al'}=2$ is
 \eqna{
&\Phi_0 = \varphi - (\pd \rho - \pd H_C)B + i (\pd H_C -3 \pd \rho)C, \\
&\Phi_1 = \sqrt{8} e^{i \rho} \bar{\psi}_j \pd x^j \tilde{A}, \\
&\Phi_{-1}=\sqrt{8} e^{-i \rho} \psi^j \pd \bar{x}_j A, \\
&\varphi  \equiv d^{\al}_{-1} W_{1\al} -  \dbar_{\ad -1} \bar{W}^{\ad}_2 + i\Pi_{-1}^m V_m +  \pd \theta^\al_{-1} V_{1\al} - \pd \bar{\theta}_{\ad -1}\bar{V}^{\ad}_2,
}[vertex]
where $\varphi$ is a superfield annihilated by modes $> 1$ and is a general linear combination of the four-dimensional supersymmetric worldsheet variables of conformal weight $+1$. Although $\Phi_1$ and $\Phi_{-1}$ do depend on the Calabi-Yau metric, we will show that they can be gauged away algebraically, so this doesn't contradict the fact that $\Phi$ is independent of the specific form of the compactification. The quantities $A$, $\tilde{A}$, $B$, $C$, $W^{\al}_1$, $\bar{W}^{\ad}_2$, $V^{\al}_1$, $\bar{V}^{\ad}_2$ and $V_m$ are usual $\mathcal{N}=1$ superfields which depend only on the zero modes of
$(x, \theta, \thetab)$, i.e., they are superfields annihilated by modes $\geq1$. The minus sign in front of $\pd \thetab_{\ad -1}$ and $\dbar_{\ad -1}$ is a consequence of the hermiticity conditions $(\pd \theta_\al)^\dagger= - \pd \bar{\theta}_{\ad}$ and $(d_{\al})^\dagger = -\dbar_{\ad}$.

\subsec{Gauge transformations}

We first look at \eqref{SFTgauge1}, \eqref{SFTgauge2} and \eqref{SFTgauge3}. In our case, $\Lambda_0=\Lambda_1=0$ and we consider
\begin{align}
&\delta \Phi_{-1} = G_6^+ \Lambda_{-1}, \\
&\delta \Phi_0 = G^+_4 \Lambda_{-1} + \tilde{G}^+_4 \Lambda_{2}, \label{gaugePhi} \\
&\delta \Phi_1  = \tilde{G}_6^+ \Lambda_2.
\end{align}
The Calabi-Yau-independent gauge parameters are
\eqna{
&\Lambda_{-1}  = \sqrt{8}e^{-i \rho}( \lambda + \pd \rho F + \pd H_C K), \\ &\Lambda_2 = \sqrt{8}e^{2i \rho - iH_C}\Big[\omega + (2 \pd \rho - \pd H_C)\tilde{F} + (3\pd \rho -2 \pd H_C) \tilde{K}\Big],
}[]
with
\begin{align*}
&\lambda  \equiv 2i (d^\al_{-1}C_{1\al} -  \dbar_{\ad -1} \bar{E}^{\ad}_2  + \pd \theta_{-1}^\al B_{1\al}  - \pd \bar{\theta}_{\ad -1} \bar{H}^{\ad}_2 + i\Pi^m_{-1}B_m), \\
&\omega  \equiv 2i (d^\al_{-1}E_{1\al} -  \dbar_{\ad -1} \bar{C}^{\ad}_2  + \pd \theta_{-1}^\al H_{1\al}  - \pd \bar{\theta}_{\ad -1} \bar{B}^{\ad}_2 + i\Pi^m_{-1}B^*_m),
\end{align*}
where $\omega$ and $\lambda$ are annihilated by modes $>1$ and $F, K, C_{1 \al}, \bar{C}^{\ad}_2, B_{ 1\al}, \bar{B}^{\ad}_2, E_{1 \al}, \bar{E}^{\ad}_2, H_{1\al}, \bar{H}^{\ad}_2$ and $B_m$ are superfields depending only on zero modes. 

Calculating the gauge transformations, we find
\eqna{
&\delta \varphi = \frac{1}{2}d_0^2 \lambda - \frac{1}{2} \dbar_0^2 \omega + \frac{1}{2} d_{-1}^2 F + \frac{1}{2} \dbar_{-1}^2 \tilde{F}, \\
&\delta iB = \frac{1}{4} \dbar^2_1 \omega - \frac{1}{4} d_1^2 \lambda + \frac{i}{2} \dbar_0^2 \tilde{F} + \frac{i}{2} d_0^2 F + \frac{3i}{4} \dbar_0^2 \tilde{K} + \frac{3i}{4} d_0^2 K, \\
&\delta C = - \frac{1}{4} \dbar_1^2 \omega - \frac{1}{4} d_1^2 \lambda - \frac{i}{2} \dbar_0^2 \tilde{F} + \frac{i}{2} d_0^2 F - \frac{i}{4} \dbar_0^2 \tilde{K} + \frac{i}{4} d_0^2 K,  \\
&\delta A =  -iK, \\
&\delta \tilde{A} = i \tilde{K}.
}[gaugetransfmodes]

From \eqref{gaugetransfmodes}, we see that the superfields appearing in $\Phi_1$ and $\Phi_{-1}$ can be gauged away algebraically using the $K$ and $\tilde{K}$ gauge parameters, as we anticipated in section \ref{sectionSFT}. Some useful relations to check gauge invariance that are obvious from \eqref{gaugetransfmodes} are
\eqna{
&\delta_\lambda iB =  \delta_\lambda C, \quad  \delta_\omega iB = -\delta_\omega C, \\
& \delta_F iB =  \delta_F C, \quad \delta_{\tilde{F}} iB = -\delta_{\tilde{F}}C, \\
&\delta_K iB = 3 \delta_K C, \quad \delta_{\tilde{K}}iB = -3 \delta_{\tilde{K}} C.
}

Following \cite{Berkovits:1998ua}, we focus on a subset of the gauge parameters. The reason for this is that $C_{1\al}$, $\bar{C}^{\ad}_2$, $B_m$, $F$, $H_{1\al}$ and $\bar{H}^{\ad}_2$ can be ignored, being parameters of $\Lambda_{-1}$ and $\Lambda_{2}$ that can be obtained from a state exact in $G^+$/$\tilde{G}^+$. After using the explicit form of $\varphi$, $\lambda$ and $\omega$ with the commutation relations for the supersymmetric modes, we obtain
  \begin{align}
&\delta W_{1\al} = 2i\Delta_{\al}^{\;\bt} B_{1\bt} + 4 \Pi_{\al \ad 0} \bar{E}^{\ad}_2 - i\dbar_0^2 E_{1\al}, \label{gaugetransfsuperfieldsW} \\
& \delta \bar{W}^{\ad}_2= -2i \bar{\Delta}^{\ad}_{\; \bd} \bar{B}^{\bd}_2 -4 \Pi^{\ad \al}_0E_{1\al} + id_0^2 \bar{E}^{\ad}_2, \label{gaugetransfsuperfieldsWbar} \\
& \delta V^m = -4i \sigma^m_{ \al \ad} \dbar_0^{\ad} E^{\al}_1  -4i \sigma^m_{\al \ad} d_0^{\al} \bar{E}^{\ad}_2, \\
& \delta \bar{V}^{\ad}_2 = - i\dbar_0^2 \bar{B}^{\ad}_2 - 16 i\bar{E}^{\ad}_2, \\
&\delta V_{1\al}= id_0^2 B_{1\al} +16iE_{1\al}, \\
& \delta (iB-C) =  \dbar_{\ad0} ( \delta \bar{W}^{\ad}_{2} - i d_0^2 \bar{E}^{\ad}_2),\\
& \delta (iB+C) =  - d^{\al}_{0} ( \delta W_{1\al} +i \dbar_0^2 E_{1\al}),
\end{align}
where $\Delta_{\al}^{\; \bt} = \{d_{\al 1}, \pd \theta^{\bt}_{-1}\} = \delta^{\bt}_{\al} + \frac{i}{2}  (\varepsilon \cdot \sigma)^{\;\;\bt}_{\al}$ and $\bar{\Delta}^{\ad}_{\;\bd}=\{\dbar^{\ad}_{1},\pd \bar{\theta}_{\bd -1} \} = \delta^{\ad}_{\bd} + \frac{i}{2} (\varepsilon \cdot \sigmabar)^{\ad}_{\;\;\bd}$.

Equations \eqref{gaugetransfsuperfieldsW} and \eqref{gaugetransfsuperfieldsWbar} imply that the superfields $W^{\al}_1$ ($\bar{W}^{\ad}_2$) can be algebraically gauge fixed to zero by choosing an appropriate $B^{\al}_1$ ($\bar{B}^{\ad}_2$), therefore, we can consistently take $W^{\al}_1=\bar{W}^{\ad}_2=0$ in the action. Imposing $W^{\al}_1 = \bar{W}^{\ad}_2 = \delta W^{\al}_1 = \delta \bar{W}^{\ad}_2 = 0$, the gauge transformations become
\begin{align}
&\delta V^m = -4i \sigma^m_{\al \ad} \dbar_0^{\ad} E_1^{\al} - 4i \sigma^m_{\al \ad} d_0^{\al} \bar{E}^{\ad}_2, \\
& \bar{\Delta}^{\ad}_{\; \bd} \delta \bar{V}^{\bd}_2 = 2 \dbar_0^2 \Pi^{\ad \al}_0 E_{1 \al} - \frac{i}{2} \dbar_0^2 d_0^2 \bar{E}^{\ad}_2 - 16i \bar{\Delta}^{\ad}_{\; \bd} \bar{E}_2^{\bd}, \\
&\Delta_{\al}^{\; \bt} \delta V_{1 \bt} = -2 d_0^2 \Pi_{\al \ad 0} \bar{E}^{\ad}_2 + \frac{i}{2} d_0^2 \dbar_0^2 E_{1 \al} + 16 i \Delta_{\al}^{\; \bt} E_{1\bt}, \\
& i \delta B = - \frac{i}{2} \Big( \dbar_{\ad 0} d_0^2 \bar{E}_2^{\ad} + d^{\al}_0 \dbar_0^2 E_{1 \al}   \Big), \\
& \delta C = \frac{i}{2} \Big( \dbar_{\ad 0} d_0^2 \bar{E}^{\ad}_2 - d_0^{\al} \dbar_0^2 E_{1 \al}  \Big).
\end{align}

\subsec{Equations of motion and superspace action}
 
Equations \eqref{SFTeom1.1} and \eqref{SFTeom1.3} give
\eqna{
& \dbar_0^2 d_0^2 A + 2 \dbar_0^2 (iB - C) = 0, \\
& d_0^2 \dbar_0^2 \tilde{A} -2 d_0^2(iB+C)=0.
}[BCcomm]

Equation \eqref{SFTeom} is more complicated to evaluate, it implies that
\begin{gather}
(d_{-1}^2 \dbar_1^2 -2 d_0^2 \dbar_0^2 + d_1^2 \dbar_{-1}^2) \varphi  \nonumber \\
+ (\dbar_0^2 d_{-1}^2  -2\dbar_{-1}^2 d_0^2 + d_0^2 \dbar_{-1}^2 -2 d_{-1}^2 \dbar_0^2)iB  \label{eom1} \\
-(-3\dbar_0^2 d_{-1}^2  +6 \dbar_{-1}^2 d_0^2 + 3 d_0^2 \dbar_{-1}^2 -6 d_{-1}^2 \dbar_0^2)C=0, \nonumber \\ \nonumber \\
(-d_0^2 \dbar_1^2 - d_1^2 \dbar_0^2 + d_2^2 \dbar_{-1}^2) \varphi \nonumber  \\ 
+ (\{d_0^2, \dbar_0^2 \} - 64)iB  \label{eom2} \\
+(-3[d_0^2, \dbar_0^2]C) - 48(d_0^2A -\dbar_0^2 \tilde{A})=0, \nonumber \\  \nonumber \\ 
(3 d_0^2 \dbar_1^2 -3 d_1^2 \dbar_0^2 + d_2^2 \dbar_{-1}^2) \varphi  \nonumber \\
-[d_0^2,  \dbar_0^2]iB \label{eom3} \\
-(64i d^{\al}_0 \Pi_{\al \ad 0} \dbar^{\ad}_0 - 19d_0^2 \dbar_0^2 -3 \dbar_0^2 d_0^2 -64)C \nonumber \\
 + 16(\dbar_0^2 \tilde{A} + d_0^2 A) = 0,\nonumber
\end{gather}
where, after acting with the generators on the string field $\Phi_0$, one obtains terms proportional to $J^{++}$, $ J^{++}(\pd \rho - \pd H_C)$ and $J^{++}(\pd H_C - 3 \pd \rho) $ which correspond respectively to the three equations above. This form of the equations is particularly useful to check gauge invariance using \eqref{gaugetransfmodes}. Some helpful relations between the modes can be found in appendix \ref{spinorconventions}.

The task now is to eliminate operators with mode numbers $\geq 1$ by using \eqref{commback}. Equations \eqref{eom2} and \eqref{eom3} can be expressed entirely in terms of zero modes (the overall oscillator number is zero), so that they only give one independent relation.  On the other hand, each term in equation \eqref{eom1} has overall oscillator mode $-1$, consequently, \eqref{eom1} will give us one independent equation for each of the supersymmetric modes $\pd \theta_{-1}^{\al}$, $\Pi^m_{-1}$, $d^{\al}_{-1}$, etc. Note that in the gauge $W^{\al}_1 = 0$, the terms proportional to $\pd \theta^{\al}_{-1}$ do not contribute to the superspace action. The same holds for terms proportional to $\pd \bar{\theta}_{\ad -1}$. Of course, to evaluate the CFT correlator corresponding to the superspace action, one needs to consider the equations of motion together with the appropriate factors of $J^{++}$, $ J^{++}(\pd \rho - \pd H_C)$ and $J^{++}(\pd H_C - 3 \pd \rho) $.

From now on,  to proceed in the computation of the action, we take $A=\tilde{A}=W^{\al}_1= \bar{W}^{\ad}_2 =0$. In this gauge,
\begin{align}
&d_{\al 1} \varphi = -2 (\sigma^m \pd \bar{\theta}_0)_{\al} V_m + \Delta_{\al}^{\;\; \bt} V_{1\bt}, \\
&\dbar^{\ad}_1 \varphi = -2 (\sigmabar^m \pd \theta_0)^{\ad} V_m - \bar{\Delta}^{\ad}_{\;\; \bd} \bar{V}^{\bd}_2, \\
&\Pi_{\al \ad 1} \varphi = -i \sigma^n_{\al \ad} ( \eta_{nm} - i \varepsilon_{nm})V^m, \\
&\pd \theta_{\al 1} \varphi = \pd \bar{\theta}^{\ad}_1 \varphi = 0.
\end{align}
For simplicity, we also perform the redefinitions
\begin{align}
&-2(\sigma^m \pd \bar{\theta}_0)_{\al} V_m + \Delta_{\al}^{\;\; \bt} V_{1 \bt} \rightarrow U_{1 \al}, \\
& 2(\sigmabar^m \pd \theta_0)^{\ad} V_m + \bar{\Delta}^{\ad}_{\;\; \bd} \bar{V}^{\bd}_2 \rightarrow \bar{U}_2^{\ad}.
\end{align}
The gauge transformations for $U_{1 \al}$ and $\bar{U}^{\ad}_2$ are then
\begin{align}
&\delta U_{1 \al} = -16i \pd \bar{\theta}^{\ad}_0 (\dbar_{\ad 0} E_{1 \al} + d_{\al 0} \bar{E}_{2 \ad}) -2 d_0^2 \Pi_{\al \ad 0} \bar{E}^{\ad}_2 + \frac{i}{2} d_0^2 \dbar_0^2 E_{1 \al} + 16i \Delta_{\al}^{\;\; \bt} E_{1 \bt},  \label{gaugeU1}\\
& \delta \bar{U}^{\ad}_2 = 16i \pd \theta_{\al 0} (\dbar^{\ad}_0 E^{\al}_1 + d^{\al}_0 \bar{E}^{\ad}_2) + 2 \dbar_0^2 \Pi^{\ad \al}_0 E_{1 \al} - \frac{i}{2} \dbar_0^2 d_0^2 \bar{E}^{\ad}_2 -16i \bar{\Delta}^{\ad}_{\;\; \bd} \bar{E}^{\bd}_2, \label{gaugeUbar2}
\end{align}
if we require that $ \delta W_1^{\al} = \delta \bar{W}_2^{\ad} =0 $ so that $W^{\al}_1$ and $ \bar{W}^{\ad}_2$ remain zero.

Computing the CFT two-point function $\bra \Phi_0^\dagger (\tilde{G}_6^+ G_6^+ + \tilde{G}_4^+ G_4^+)\Phi_0 \ket$, one finds that the string field theory action in $\mathcal{N}=1$ superspace for the first massive compactification-independent fields of the charged open superstring coupled to a constant electromagnetic background is
\eqna{
S&= \frac{1}{16} \int d^4x \; p_0^2 \bar{p}_0^2 \bigg\{  V^{\dagger}_n(\eta^{nm} - i \varepsilon^{nm})\bigg[ -\{d_0^2 , \dbar_0^2 \} V_m + 16 \Pi^n_0 \Pi_{n 0} V_m -32 (\eta_{mp} -i \varepsilon_{mp})V^p \\
&-32\Big( (\pd \bar{\theta}_0 \dbar_0)V_m + (\pd \theta_0 d_0) V_m \Big) + 8 \sigmabar^{\ad \al}_m \Big( d_{\al 0} \bar{U}_{2 \ad} - \dbar_{\ad 0}U_{1 \al} \Big) + 32 \Pi_{m 0} B \\
&+ 24 \sigmabar^{\ad \al}_m [ \dbar_{\ad 0}, d_{\al 0}]C \bigg] + U^{\al}_2 \bigg[ -8 \sigma^n_{\al \ad} (\eta_{nm} - i\varepsilon_{nm}) \dbar^{\ad}_0 V^m + 4 \dbar_{\ad 0} d_{\al 0} \bar{U}^{\ad}_2 - 4 \dbar_0^2 U_{1 \al} \\
& + d_{\al 0} \dbar_0^2 (-2iB + 18C) + \pd \theta_{\al 0} (-32iB - 96C) - 48i \Pi_{\al \ad 0} \dbar^{\ad}_0 C \bigg] \\
& - \bar{U}_{1 \ad} \bigg[ -8 \sigmabar^{n \ad \al} (\eta_{nm} -i \varepsilon_{nm}) d_{\al 0} V^m + 4 d_0^2 \bar{U}^{\ad}_2 - 4 d^{\al}_0 \dbar^{\ad}_0 U_{1 \al} - \dbar^{\ad}_0 d_0^2 (2iB + 18C) \\
&+ \pd \bar{\theta}^{\ad}_0 (-32i B + 96C) + 48 i \Pi^{\ad \al}_0 d_{\al 0} C \bigg] + B^{\dagger} \bigg[ -32 \Pi^n_0 (\eta_{nm} - i\varepsilon_{nm}) V^m \\
& + (\{ d_0^2 , \dbar_0^2 \} - 64)B + 3i [d_0^2, \dbar_0^2]C - i \Big( 2 d_0^2\dbar_{\ad 0} + 32 \pd \bar{\theta}_{\ad 0} \Big) \bar{U}^{\ad}_2 + i \Big( 2 \dbar_0^2 d^{\al}_0 + 32 \pd \theta^{\al}_0\Big) U_{1 \al} \bigg] \\
& + 3C^{\dagger} \bigg[ -8 \sigmabar^{n \ad \al} [d_{\al 0}, \dbar_{\ad 0}] (\eta_{nm} - i \varepsilon_{nm}) V^m - \Big( 6 d^{\al}_0 \dbar_0^2 + 8i \Pi^{\ad \al}_0 \dbar_{\ad 0} \Big) U_{1 \al} \\
& - \Big( 6 \dbar_{\ad 0} d_0^2 + 8i \Pi_{\al \ad 0} d^{\al}_0 \Big) \bar{U}^{\ad}_2 - [d_0^2, \dbar_0^2]i B \\
& - \big( -11 \{ d_0^2, \dbar_0^2 \} + 128 \Pi^n_0 \Pi_{n 0} -256 \pd \bar{\theta}_{\ad 0} \dbar^{\ad}_0 - 256 \pd \theta^{\al}_0 d_{\al 0} -64 \big) C \bigg] \bigg\},
}[actionmassiveback]
with the equations of motion \begingroup
\allowdisplaybreaks
\begin{align}
& - \{d_0^2, \dbar_0^2 \} V_m + 16 \Pi^n_0 \Pi_{n 0} V_m -32 ( \eta_{mn} - i\varepsilon_{mn}) V^n -32 \Big[ (\pd \bar{\theta}_0 \dbar_0)V_m + ( \pd \theta_0 d_0)V_m \Big] \nonumber \\
& + 8 \sigmabar^{\ad \al}_m ( d_{\al 0} \bar{U}_{2 \ad} - \dbar_{\ad 0} U_{1 \al} ) + 32 \Pi_{m 0} B + 24 \sigma_{m \al \ad} [\dbar^{\ad}_0, d^{\al}_0]C = 0,  \label{eomback1}  \\
\nonumber  \\
&  -8 \sigma^n_{\al \ad} (\eta_{nm} - i\varepsilon_{nm}) \dbar^{\ad}_0 V^m +4 \dbar_{\ad 0} d_{\al 0} \bar{U}^{\ad}_2 - 4 \dbar_0^2 U_{1 \al} + d_{\al 0} \dbar_0^2 (-2i B + 18C) \nonumber \\
&+ \pd \theta_{\al 0} ( -32iB - 96C) -48i \Pi_{\al \ad 0} \dbar^{\ad}_0 C = 0, \label{eomback2}\\
\nonumber  \\
& -8 \sigmabar^{n \ad \al} ( \eta_{nm} - i \varepsilon_{nm}) d_{\al 0} V^m + 4 d_0^2 \bar{U}^{\ad}_2  -4 d^{\al}_0 \dbar^{\ad}_0 U_{1 \al} - \dbar^{\ad}_0 d_0^2 ( 2iB +18C) \nonumber \\
& + \pd \bar{\theta}^{\ad}_0 (-32iB + 96C) + 48i \Pi^{\ad \al}_0 d_{\al 0} C = 0, \label{eomback3}\\
\nonumber \\
\nonumber \\
& -32i \Pi^n_0 (\eta_{nm} - i \varepsilon_{nm})V^m + (\{d_0^2, \dbar_0^2 \} -64) iB - 3 [d_0^2, \dbar_0^2]C \nonumber  \\
& + (2 d_0^2 \dbar_{\ad 0} + 32 \pd \bar{\theta}_{\ad 0}) \bar{U}^{\ad}_2 - (2\dbar_0^2 d^{\al}_0 + 32 \pd \theta^{\al}_0) U_{1 \al} = 0, \label{eomback4} \\
\nonumber  \\
&  -8 \sigmabar^{n \ad \al} [ d_{\al 0}, \dbar_{\ad 0}] (\eta_{nm} - i \varepsilon_{nm})V^m \nonumber  \\
& - \left( 6 d^{\al}_0 \dbar_0^2 + 8i \Pi^{\ad \al}_0 \dbar_{\ad 0} \right) U_{1 \al} - \left( 6 \dbar_{\ad 0} d_0^2 + 8i \Pi_{\al \ad 0} d^{\al}_0 \right) \bar{U}^{\ad}_2 \nonumber  \\
&- [d_0^2, \dbar_0^2] iB - \left( -11\{d_0^2, \dbar_0^2 \} + 128 \Pi^n_0 \Pi_{n 0} -256 \pd \bar{\theta}_{\ad 0} \dbar^{\ad}_0 -256 \pd \theta^{\al}_0 d_{\al 0} -64 \right) C = 0. \label{eomback5}
\end{align}%
\endgroup

Using \eqref{eomback2}, \eqref{eomback3}, \eqref{eomback4}, \eqref{eomback5} and \eqref{BCcomm}, one can show that
\begin{equation}
B   = - \frac{i}{32} \bigg[ [d_0^2, \dbar_0^2] C + \dbar_{\ad} d_0^2 \bar{U}^{\ad}_2 - d^{\al}_0 \dbar_0^2 U_{1 \al} \bigg]  + \frac{i}{2} (\pd \theta_0 d_0) (3C + iB) - \frac{i}{2} (\pd \bar{\theta}_0 \dbar_0) (3C -iB), \label{relationforB}
\end{equation}

and
\eqna{
 64( \Pi^n_0 \Pi_{n 0} + 1 )C &= 6 \{d_0^2 , \dbar_0^2 \}C -2 \dbar_{\ad 0} d_0^2  \bar{U}^{\ad}_2 -2 d^{\al}_0 \dbar_0^2 U_{1 \al} \\
& + 32 (\pd \theta_0 d_0)(-iB + C) +32 (\pd \bar{\theta}_0 \dbar_0) (iB + C). 
}[relationforC]
Equations \eqref{relationforB} and \eqref{relationforC} generalize (3.8) and (3.9) from \cite{Berkovits:1998ua} for the uncharged and non-interacting case. At the level of the equations of motion, one can also gauge-fix $U_{1 \al} = \bar{U}_2^{\ad} = 0$ by the gauge transformations \eqref{gaugeU1} and \eqref{gaugeUbar2}. Note that using our conventions for the supersymmetric variables, the quadratic action of \cite{Berkovits:1998ua} for the non-interacting case is 

\begin{equation}
\begin{split}
S_{free} &= \frac{1}{16} \int d^4x \; p_0^2 \bar{p}_0^2 \bigg\{ V^m \bigg[ -\{ d_0^2, \dbar_0^2\}V_m + 16 \Pi^n_0 \Pi_{n 0} V_m -32 V_m \\
&+ 16 \sigmabar_m^{\ad \al}(d_{\al 0} \bar{V}_{\ad} - \dbar_{\ad 0} V_{\al})  + 64 \Pi_{m0} B + 48 \sigmabar_m^{\ad \al} [\dbar_{\ad 0}, d_{\al 0}]C \bigg] \\
& + V^{\al} \bigg[ 8 \dbar_{\ad 0} d_{\al 0} \bar{V}^{\ad} - 4 \dbar_0^2 V_{\al} + 2 d_{\al 0} \dbar_0^2 (-2iB + 18C) -96i \Pi_{\al \ad 0} \dbar^{\ad}_0 C \bigg] \\
& + \bar{V}_{\ad} \bigg[ -4 d_0^2 \bar{V}^{\ad} + 2 \dbar^{\ad}_0 d_0^2 ( 2iB +18C) -96i \Pi^{\ad \al}_0 d_{\al 0} C \bigg] \\
& + B \bigg[ \{d_0^2, \dbar_0^2 \}B -64B + 6i [d_0^2, \dbar_0^2]C \bigg]  + 3C \bigg[ 11\{ d_0^2, \dbar_0^2 \} C -128 \Pi^n_0 \Pi_{n 0} C + 64C \bigg] \bigg\}. \label{actionfreecasenathanpaper}
\end{split}
\end{equation}

\newsec{Equations of Motion in Lorenz Gauge }[secsix]

Although one can in principle try to analyze directly the equations of motion of \eqref{eomback1} -  \eqref{eomback5}, it will be much simpler to first choose a convenient gauge.
In bosonic open string field theory, the equation of motion $\mathcal{Q}V =0$ together with the gauge transformation $\delta V = \mathcal{Q} \Lambda$ implies that one can choose a gauge in which $V$ is a conformal
primary field of weight zero, i.e., it has no double or higher-order poles with the stress tensor $T$.
For the massless vector field $A_m$, this correponds to Lorenz gauge $\partial_m A^m =0$. Similarly, in the hybrid formalism, the equations of motion and gauge invariances imply that one can choose a Lorenz-type gauge in which the string field is an $\mathcal{N}=2$ superconformal primary field of weight zero,
i.e., it has no double poles with the $\mathcal{N}=2$ superconformal generators $(T, G^+, G^-, J)$.

In this section, we impose these $\mathcal{N}=2$ superconformal primary conditions on the string field $\Phi_0$ and compute the equations of motion in this gauge for the component fields of the first massive state of the charged open superstring coupled to a constant electromagnetic background in four dimensions. These are the on-shell relations implied by the superspace action \eqref{actionmassiveback}, which describes 12 complex bosonic and 12 complex fermionic degrees of freedom on-shell.

\subsec{String field and physical conditions}

For a string field $\Phi_0$ to be physical with zero $U(1)$ charge and zero conformal weight, it should have no double poles or higher in the OPE with $(G^{\pm}, T)$ and no single poles or higher in the OPE with $J$. Demanding these conditions to be satisfied, we will obtain the gauge-fixed equations of motion for the physical degrees of freedom.

We will work in the gauge $W^{\al}_1 = \bar{W}^{\ad}_2 = 0$, so that our string field is
\eqn{
\Phi_0 = i\Pi_{-1}^m V_m +  \pd \theta^\al_{-1} V_{1\al} - \pd \bar{\theta}_{\ad -1}\bar{V}^{\ad}_2  - (\pd \rho - \pd H_C)B + i (\pd H_C -3 \pd \rho)C.
}[]
Note that, in \cite{Berkovits:1997zd}, the string field was incomplete because the authors didn't include terms proportional do $\pd H_C$. This is equivalent to putting the superfield $C$ equal to zero and considering only the massive spin-2 multiplet. Nevertheless, the results in \cite{Berkovits:1997zd} are correct for the free case, because of a field redefinition that we shall present shortly which decouples the superfield $C$.

The vanishing of the double poles in the OPE of $J$ and $\Phi_0$ imply the condition 
\eqn{
B=0.
}[]
Given that the superfields only depend on the four-dimensional coordinates $x^m$, only $(G^{\pm}_4, T)$ will produce double poles with $\Phi_0$. The condition that the double poles in the OPE of $\Phi_0$ with $G^{\pm}_4$ vanish results in
\begin{equation*}
\begin{split}
& d_0^2 C = 0, \\
&d_0^2 V_{1\al} = 0,  \\
&d_0^2 V_m=0,  \\ 
&2 \Delta_{\al}^{\;\bt} V_{1\bt}  - 4 \sigma^m_{\al \ad} \pd \bar{\theta}^{\ad}_0 V_m -6 d_{\al 0} C = 0,  \\
&d_0^2 \bar{V}^{\ad}_2 - 4 \sigmabar^{m \ad \al} d_{ \al 0} V_m = 0,
\end{split}
\quad
\begin{split}
& \dbar^2_0 C = 0, \\
&\dbar_0^2 \bar{V}^{\ad}_{2} = 0, \\
&\dbar_0^2 V_m = 0, \\
&2 \bar{\Delta}^{\ad}_{\;\bd} \bar{V}^{\bd}_2 + 4 \sigmabar^{\ad \al}_m \pd \theta_{\al 0} V^m -6 \dbar^{\ad}_0 C = 0, \\
&\dbar_0^2 V_{1\al} + 4 \sigma^m_{\al \ad} \dbar_0^{\ad} V_m = 0.
\end{split}
\end{equation*}

The equations above enable us to eliminate the superfields $V^{\al}_1$, $\bar{V}^{\ad}_2$ and, after redefining $V_m \rightarrow (1- i \varepsilon)_{mn} V^n$ and $C \rightarrow \frac{4}{3} C$, we obtain the following set of independent equations
\begin{align}
&d_0^2 C = 0,  \label{eomCLorenz1}\\
&\dbar_0^2 C = 0, \label{eomCLorenz2}\\
&\sigmabar^{m \ad \al} d_{\al 0} V_m - d_0^2 \dbar^{\ad}_0 C =  0, \label{eomVmLorenz1} \\
& \sigma^m_{\al \ad} \dbar^{\ad}_0 V_m + \dbar^2_0 d_{\al 0} C = 0. \label{eomVmLorenz2}
\end{align}
Note that these equations imply $\dbar_0^2 V_m = d_0^2 V_m = 0$. 

In components, we write the complex superfields as
\begin{align}
V_m &= 4C_m + i(\theta_0 \chi_{1m}) - i (\bar{\theta}_0 \bar{\chi}_{2m}) + i (\theta_0 \theta_0)(M_{1m}+iN_{1m}) -i (\bar{\theta}_0 \bar{\theta}_0)(M_{2m} -i N_{2m}) \nonumber \\
& + (\theta_0 \sigma^n \bar{\theta}_0) v_{mn} + \frac{i}{\sqrt{2}} (\theta_0 \theta_0) (\bar{\theta}_0 \bar{\lambda}_{1m})  - \frac{i}{\sqrt{2}} (\bar{\theta}_0 \bar{\theta}_0) (\theta_0 \lambda_{2m}) + (\theta_0 \theta_0) (\bar{\theta}_0 \bar{\theta}_0) D_m, \label{superfieldVm} \\
 C &= \phi + i(\theta_0 \xi_1) - i (\bar{\theta}_0 \bar{\xi}_{2})  + i (\theta_0 \theta_0)(M_{1}+iN_{1}) -i (\bar{\theta}_0 \bar{\theta}_0)(M_{2} -i N_{2}) \nonumber \\
 &+ (\theta_0 \sigma^n \bar{\theta}_0) a_n + \frac{i}{\sqrt{2}} (\theta_0 \theta_0) (\bar{\theta}_0 \bar{\psi}_{1})
 - \frac{i}{\sqrt{2}}(\bar{\theta}_0 \bar{\theta}_0) (\theta_0 \psi_{2}) + (\theta_0 \theta_0) (\bar{\theta}_0 \bar{\theta}_0) D.
\end{align}
As will be explicitly shown, the fermionic sector of the complex superfield $V_m$ describes two massive spin-3/2 Dirac fermions with components $ (\chi_{1m} , \bar{\lambda}_{1m})$ and $ (\chi_{2m}, \bar{\lambda}_{2m})$, while for C it describes two massive spin-1/2 Dirac fermions $(\xi_1, \bar{\psi}_1)$ and $(\xi_2, \bar{\psi}_2)$. For the bosonic sector, $V_m$ describes a massive complex vector $C_m$ and a massive complex spin-2 field $v_{mn}$, while $C$ describes a massive complex scalar $\phi$ and a massive complex vector $a_m$. The remaining fields are auxiliary.

\subsec{Fermionic components}

The vanishing of the double poles of $\Phi_0$ with $T$ imply the mass-shell conditions
\begin{align}
&( \mathfrak{D}^2 -2) \xi_{1 \al} - i \varepsilon_{rs} (\sigma^{rs} \xi_1)_{\al} = 0, \label{mass_xi} \\
&( \mathfrak{D}^2 -2) \bar{\psi}^{\ad}_1 - i \varepsilon_{rs} (\sigmabar^{rs} \bar{\psi}_1)^{\ad} = 0, \label{mass_psi} \\
& (\mathfrak{D}^2 -2) \chi_{1\al m} + 2i \varepsilon_{mn} \chi^n_{1\al} - i \varepsilon_{rs} (\sigma^{rs} \chi_{1m})_{\al} = 0, \label{mass_chi} \\
& ( \mathfrak{D}^2 -2) \bar{\lambda}^{\ad}_{1m} + 2i \varepsilon_{mn} \bar{\lambda}^{\ad n}_1 - i \varepsilon_{rs} (\sigmabar^{rs} \bar{\lambda}_{1m})^{\ad} = 0, \label{mass_lambda}
\end{align}
and similar relations hold respectively for $\psi_{2 \al}$, $\bar{\xi}_{2}^{\ad}$, $\lambda_{2 \al m}$ and $\bar{\chi}_{2 m}^{\ad}$.

From equations \eqref{eomCLorenz1} and \eqref{eomCLorenz2}, one has
\begin{align}
&i \mathfrak{D}_n (\sigmabar^n \xi_1)^{\ad} + \sqrt{2} \bar{\psi}^{\ad}_1 = 0, \\
&i \mathfrak{D}_n (\sigma^n \bar{\xi}_2)_{\al} + \sqrt{2} \psi_{2 \al}= 0, \\
& \sqrt{2}i \mathfrak{D}_n (\sigma^n \bar{\psi}_1)_{\al} + \mathfrak{D}^2 \xi_{1 \al} - i \varepsilon_{rs} (\sigma^{rs} \xi_{1 })_{\al} = 0, \\
& \sqrt{2} i \mathfrak{D}_n (\sigmabar^n \psi_2)^{\ad} + \mathfrak{D}^2 \bar{\xi}^{\ad}_2 - i \varepsilon_{rs} (\sigmabar^{rs} \bar{\xi}_2)^{\ad} = 0.
\end{align}
 Together with the mass-shell conditions, these give the remaining equations of motion for the massive spin-1/2
 \begin{align}
 &i \mathfrak{D}_n (\sigma^n \bar{\psi}_1)_{\al} + \sqrt{2} \xi_{1 \al} = 0, \\
 & i \mathfrak{D}_n (\sigmabar^n \psi_2)^{\ad} + \sqrt{2} \bar{\xi}^{\ad}_2 =0.
 \end{align}

 The shortest path to obtain the equations of motion for the massive spin-3/2 is similar to what we did for the spin-1/2 case. First, one calculates $d^2_0 V_m = \dbar^2_0 V_m = 0$ in components, and then uses the mass-shell equations to get
 \begin{align}
 &i\mathfrak{D}_n (\sigmabar^n \chi_{1m})^{\ad} + \sqrt{2} \bar{\lambda}_{1m}^{\ad} =0,  \label{eom_chi1_back}\\
&i \mathfrak{D}_n (\sigma^n \bar{\chi}_{2m})_{\al} + \sqrt{2} \lambda_{2 \al m} = 0, \label{eom_chi2_back} \\
&i\mathfrak{D}_n (\sigma^n \bar{\lambda}_{1m})_{\al} + \sqrt{2} \chi_{1\al m} - i \sqrt{2} \varepsilon_{mn} \chi^n_{1\al} = 0, \\
& i\mathfrak{D}_{n} ( \sigmabar^n \lambda_{2m})^{\ad} + \sqrt{2} \bar{\chi}_{2m}^{\ad} - i \sqrt{2} \varepsilon_{mn} \bar{\chi}^{\ad n}_2 = 0. \label{eom_lambda2_back}
 \end{align}

The generalization of the Lorenz conditions obeyed by the spin-3/2 fields can be calculated by summing the expressions obtained by acting with $\dbar_{\ad 0}$ in \eqref{eomVmLorenz1}  and $d^{\al}_0$ in \eqref{eomVmLorenz2}. One finds
\begin{align}
&\mathfrak{D}^m \chi_{1 \al m} - 2 \varepsilon_{rs} (\sigma^{rs} \xi_1)_{\al} = 0, \\
& \mathfrak{D}^m \bar{\chi}^{\ad}_{2m} + 2 \varepsilon_{rs} (\sigmabar^{rs} \bar{\xi}_2)^{\ad} = 0, \\
& \mathfrak{D}^m \bar{\lambda}^{\ad}_{1m} -i \frac{\sqrt{2}}{2} \tilde{\varepsilon}^{mn} (\sigmabar_n \chi_{1m})^{\ad} + 2 \varepsilon_{rs} (\sigmabar^{rs} \bar{\psi}_1)^{\ad} - 2 \sqrt{2} i \varepsilon^{mn} \mathfrak{D}_m (\sigmabar_n \xi_1)^{\ad} = 0, \\
& \mathfrak{D}^m \lambda_{2 \al m} + i \frac{\sqrt{2}}{2} \tilde{\varepsilon}^{mn} (\sigma_n \bar{\chi}_{2m})_{\al} - 2 \varepsilon_{rs} (\sigma^{rs} \psi_2)_{\al} + 2 \sqrt{2} i \varepsilon^{mn} \mathfrak{D}_m (\sigma_n \bar{\xi}_2)_{\al} = 0,
\end{align}
where $\tilde{\varepsilon}^{mn} = \frac{1}{2} \epsilon^{mnrs} \varepsilon_{rs}$ and $\epsilon_{0123} = -1$.

Contracting equations \eqref{eom_chi1_back} and \eqref{eom_chi2_back} with a sigma matrix in the Lorentz vector index and using the Lorenz conditions, we obtain the sigma trace constraints for $\bar{\lambda}_{1m}^{\ad}$ and $\lambda_{2\al m}$. The trace constraints for $\chi_{1 \al m}$ and $\bar{\chi}_{2m}^{\ad}$ come from evaluating the $(\theta_0)^0 (\bar{\theta}_0)^0$ term in equations \eqref{eomVmLorenz1} and \eqref{eomVmLorenz2}. And, after using the equations of motion for the spin-1/2 fields, these constraints  can be expressed as
\begin{align}
&(\sigmabar^m \chi_{1m})^{\ad} + 4 \sqrt{2} \bar{\psi}_1^{\ad} = 0, \label{tracelessLorenz1} \\
& (\sigma^m \bar{\chi}_{2m})_{\al} - 4 \sqrt{2} \psi_{2 \al} = 0, \\
&(\sigma^m \bar{\lambda}_{1m})_{\al} - 4 \sqrt{2} \xi_{1 \al} - 2 \sqrt{2} i \varepsilon_{rs} (\sigma^{rs} \xi_1)_{\al} = 0, \\
&(\sigmabar^m \lambda_{2m})^{\ad} + 4 \sqrt{2} \bar{\xi}_2^{\ad} + 2 \sqrt{2} i \varepsilon_{rs} (\sigmabar^{rs} \bar{\xi}_2)^{\ad} = 0. \label{tracelessLorenz2}
\end{align}

The generators of Lorentz transformations for the vector and spin-1/2 representations are $(J^{rs})_{mp} = -i (\eta^r_m \eta^s_p - \eta^s_m \eta^r_p) $ and $ S^{rs} = \text{diag}(i\sigma^{rs}, i\sigmabar^{rs})$, so it is straightforward to see that equations \eqref{mass_xi} to \eqref{mass_lambda} describe fermions with the gyromagnetic ratio $g=2$. This is expected from the known result that all particles in string theory have $g=2$ \cite{Ferrara:1992yc}.

As a last comment in this subsection, we recall that a ubiquitous problem that arises when one tries to write equations of motion coming from an action, for a charged massive spin-3/2 field in an electromagnetic background, is the loss of causal propagation of the physical degrees of freedom \cite{Velo:1969bt}\cite{Cox:1989hp}\cite{Deser:2000dz}\cite{Porrati:2009bs}. In our case, the spin-3/2 is not alone, i.e., it comes coupled to spin-1/2 fields. However, the propagation remains causal in the presence of the background. This can be easily shown by the method of characteristics explained in \cite{Argyres:1989cu}\cite{Porrati:2010hm}\cite{Cox:1989hp}, which establishes that equations \eqref{eom_chi1_back} to \eqref{eom_lambda2_back} do not have spacelike characteristic surfaces.

\subsec{Bosonic components}

The vanishing of the double poles of $\Phi_0$ with $T$ imply the equations of motions
\begin{align}
&(\mathfrak{D}^2 -2) \phi = 0, \label{phiM_shell}\\
&(\mathfrak{D}^2 -2)a_m +2i \varepsilon_{mn} a^n = 0,\\
&(\mathfrak{D}^2 -2)C_m + 2i \varepsilon_{mn} C^n = 0,\\
& (\mathfrak{D}^2 -2) v_{mp} + 2i \Big( \varepsilon_m^{\;\;n} v_{np} - v_{mn} \varepsilon^n_{\;\;p} \Big) = 0,\label{vmnM_shell}
\end{align}
and equations \eqref{phiM_shell} to \eqref{vmnM_shell} make manifest the fact that the bosonic fields involved have a gyromagnetic ratio of 2. 

From $d_0^2 C = \dbar_0^2 C = d_0^2 V_m = \dbar_0^2 V_m =0$, one obtains
\begin{align}
&D_m = -  \mathfrak{D}^2 C_m, \\
&D = - \frac{1}{4} \mathfrak{D}^2 \phi, \\
&\mathfrak{D}^n v_{mn} = 0, \label{almostlorenzvmn}\\
& \mathfrak{D}^n a_n = 0, \\
&M_{1m} = N_{1m} = M_{2m} = N_{2m} = M_1 = N_1 = M_2 =N_2 = 0.
\end{align}

To find the remaining constraints, we calculate the terms proportional to $(\bar{\theta}_0)^1$ and $(\theta_0)^1$ of equations \eqref{eomVmLorenz1} and \eqref{eomVmLorenz2}, respectively. One finds
\begin{align}
&(\sigmabar^m \sigma^n)^{\ad}_{\;\bd} \Big( v_{mn}  + 4i \mathfrak{D}_n C_m -4i \mathfrak{D}_m a_n \Big) - 8 \delta^{\ad}_{\bd} \phi - 4i \varepsilon_{rs} \sigmabar^{rs \ad}_{\;\;\;\;\;\; \bd} \phi = 0,  \label{eqbosonconst1} \\
&(\sigma^m \sigmabar^n)_{\al}^{\; \bt} \Big( -v_{mn} + 4i \mathfrak{D}_n C_m -4i \mathfrak{D}_m a_n \Big) + 8 \delta^{\bt}_{\al} \phi + 4i \varepsilon_{rs} \sigma^{rs \;  \bt}_{\;\;\; \al} \phi = 0. \label{eqbosonconst2}
\end{align}
Taking the trace of \eqref{eqbosonconst1} and \eqref{eqbosonconst2}, one arrives at
\begin{align}
&\mathfrak{D}^m C_m = 0, \\
& v^n_{\;\; n} + 8 \phi = 0. \label{tracev_mn}
\end{align}

Contracting \eqref{eqbosonconst1} with $(\sigmabar^p \sigma^q)^{\bd}_{\; \ad}$ and \eqref{eqbosonconst2} with $(\sigma^p \sigmabar^q)_{\bt}^{\; \al}$ and taking the difference of the results, it is found that the antisymmetric part of $v_{mn}$ is not independent
\begin{equation}
v_{[mn]} = 2 \epsilon_{mnpq} \Big( \mathfrak{D}^p C^q +  \mathfrak{D}^p a^q \Big) + 2i \varepsilon_{mn} \phi.
\end{equation}
One can plug this back in \eqref{almostlorenzvmn} to find the Lorenz gauge condition for the symmetric part of $v_{mn}$
\begin{equation}
\mathfrak{D}^m v_{(mn)} = - 2i \tilde{\varepsilon}_{nm} C^m - 2i \tilde{\varepsilon}_{nm} a^m -2i \varepsilon_{nm} \mathfrak{D}^m \phi. \label{Lorenzv_mnconst}
\end{equation}
Note that \eqref{vmnM_shell} also implies the equation of motion for $v_{(mn)}$
\begin{equation}
( \mathfrak{D}^2 -2) v_{(mp)} + 2i \Big( \varepsilon_m^{\;\;n} v_{(np)} - v_{(mn)}\varepsilon^n_{\;\;p}\Big) = 0. \label{eomv_mnNappi}
\end{equation}

We mention that the equation of motion \eqref{eomv_mnNappi} for the massive spin-2 in four dimensions is identical to the one obtained from bosonic string theory \cite{Porrati:2011uu}, as expected. However, the constraints \eqref{tracev_mn} and \eqref{Lorenzv_mnconst} are not the same as in \cite{Porrati:2011uu}, for the obvious reason that the fields present in the first massive state of the superstring are different than those of the bosonic string. Consequently, the component action will take a different form in the case of the superstring.

As was mentioned earlier, in the free case, one can perform the field redefinition $V_m \rightarrow V_m + \frac{1}{3} \sigmabar_m^{\ad \al} [d_{\al 0}, \dbar_{\ad 0}]C$ in \eqref{eomVmLorenz1} and \eqref{eomVmLorenz2} to achieve the results of \cite{Berkovits:1997zd} in which $C$ is decoupled.
 In components, this field redefinition reads
\begin{align}
& C_m \rightarrow C_m - \frac{1}{3} a_m, \\
& v_{mn} \rightarrow v_{mn} - 2 \eta_{mn} \phi + \frac{4}{3} \epsilon_{mnpq} \partial^p a^q, \\
&\chi_{1 \al m} \rightarrow \chi_{1 \al m} - \frac{4}{3} \Big( i \partial_m \xi_{1 \al} - \sqrt{2} (\sigma_m \bar{\psi}_1)_{\al} \Big), \\
&\bar{\lambda}^{\ad}_{1m} \rightarrow \bar{\lambda}^{\ad}_{1m} + \frac{4}{3} \Big( i \partial_m \bar{\psi}^{\ad}_1 - \sqrt{2} (\sigmabar_m \xi_1)^{\ad} \Big), \\
&\bar{\chi}^{\ad}_{2m} \rightarrow  \bar{\chi}^{\ad}_{2m} + \frac{4}{3} \Big( i \partial_m \bar{\xi}^{\ad}_2 - \sqrt{2} (\sigmabar_m \psi_2)^{\ad} \Big), \\
&\lambda_{2 \al m} \rightarrow \lambda_{2 \al m} - \frac{4}{3} \Big( i \partial_m \psi_{2 \al} - \sqrt{2} (\sigma_m \bar{\xi}_2)_{\al} \Big).
\end{align}
 Unfortunately, we did not encounter a field redefinition that makes this simplification for the interacting case. An obvious guess would be $V_m \rightarrow V_m + A_1 \sigmabar_m^{\ad \al} [d_{\al 0}, \dbar_{\ad 0}]C + A_2 \Pi_mC$ with arbitrary coefficients $A_1$ and $A_2$. But it doesn't work as desired, the problem being the appearance of terms proportional to $\partial \theta_0^{\al}$ and $\partial \bar{\theta}_0^{\ad}$ multiplying $C$ that cannot be eliminated. One could also try to work in components and generalize the field redefinitions that worked in the free case by modifying $\pd_m \rightarrow \mathfrak{D}_m$. This change is not ideal either, for example, it generates terms coupling the background to the lower spin fields in the Dirac equations \eqref{eom_chi1_back} to \eqref{eom_lambda2_back}, so that it does not provide any simplification.

Finally, we summarize below our results for this section:

\subsection*{Massive scalar}
Equation of motion
\begin{equation}
(\mathfrak{D}^2 -2) \phi = 0.
\end{equation}

\subsection*{Massive spin-1/2}
Equations of motion
\begin{align}
&i \mathfrak{D}_n (\sigmabar^n \xi_1)^{\ad} + \sqrt{2} \bar{\psi}^{\ad}_1 = 0, \\
&i \mathfrak{D}_n (\sigma^n \bar{\xi}_2)_{\al} + \sqrt{2} \psi_{2 \al}= 0, \\
&i \mathfrak{D}_n (\sigma^n \bar{\psi}_1)_{\al} + \sqrt{2} \xi_{1 \al} = 0, \\
& i \mathfrak{D}_n (\sigmabar^n \psi_2)^{\ad} + \sqrt{2} \bar{\xi}^{\ad}_2 =0.
\end{align}

\subsection*{Massive vectors}
Equation of motion
\begin{equation}
(\mathfrak{D}^2 -2)C_m + 2i \varepsilon_{mn} C^n = 0.
\end{equation}
Lorenz condition
\begin{equation}
\mathfrak{D}^m C_m = 0.
\end{equation}
The same expressions hold for $a_m$.

\subsection*{Massive spin-3/2}
Equations of motion
\begin{align}
 &i\mathfrak{D}_n (\sigmabar^n \chi_{1m})^{\ad} + \sqrt{2} \bar{\lambda}_{1m}^{\ad} =0,  \\
&i \mathfrak{D}_n (\sigma^n \bar{\chi}_{2m})_{\al} + \sqrt{2} \lambda_{2 \al m} = 0,  \\
&i\mathfrak{D}_n (\sigma^n \bar{\lambda}_{1m})_{\al} + \sqrt{2} \chi_{1\al m} - i \sqrt{2} \varepsilon_{mn} \chi^n_{1\al} = 0, \\
& i\mathfrak{D}_{n} ( \sigmabar^n \lambda_{2m})^{\ad} + \sqrt{2} \bar{\chi}_{2m}^{\ad} - i \sqrt{2} \varepsilon_{mn} \bar{\chi}^{\ad n}_2 = 0.
 \end{align}
 Lorenz gauge conditions
\begin{align}
&\mathfrak{D}^m \chi_{1 \al m} - 2 \varepsilon_{rs} (\sigma^{rs} \xi_1)_{\al} = 0, \\
& \mathfrak{D}^m \bar{\chi}^{\ad}_{2m} + 2 \varepsilon_{rs} (\sigmabar^{rs} \bar{\xi}_2)^{\ad} = 0, \\
& \mathfrak{D}^m \bar{\lambda}^{\ad}_{1m} -i \frac{\sqrt{2}}{2} \tilde{\varepsilon}^{mn} (\sigmabar_n \chi_{1m})^{\ad} + 2 \varepsilon_{rs} (\sigmabar^{rs} \bar{\psi}_1)^{\ad} - 2 \sqrt{2} i \varepsilon^{mn} \mathfrak{D}_m (\sigmabar_n \xi_1)^{\ad} = 0, \\
& \mathfrak{D}^m \lambda_{2 \al m} + i \frac{\sqrt{2}}{2} \tilde{\varepsilon}^{mn} (\sigma_n \bar{\chi}_{2m})_{\al} - 2 \varepsilon_{rs} (\sigma^{rs} \psi_2)_{\al} + 2 \sqrt{2} i \varepsilon^{mn} \mathfrak{D}_m (\sigma_n \bar{\xi}_2)_{\al} = 0.
\end{align}
Sigma trace constraints
\begin{align}
&(\sigmabar^m \chi_{1m})^{\ad} + 4 \sqrt{2} \bar{\psi}_1^{\ad} = 0,  \\
& (\sigma^m \bar{\chi}_{2m})_{\al} - 4 \sqrt{2} \psi_{2 \al} = 0, \\
&(\sigma^m \bar{\lambda}_{1m})_{\al} - 4 \sqrt{2} \xi_{1 \al} - 2 \sqrt{2} i \varepsilon_{rs} (\sigma^{rs} \xi_1)_{\al} = 0, \\
&(\sigmabar^m \lambda_{2m})^{\ad} + 4 \sqrt{2} \bar{\xi}_2^{\ad} + 2 \sqrt{2} i \varepsilon_{rs} (\sigmabar^{rs} \bar{\xi}_2)^{\ad} = 0. 
\end{align}

\subsection*{Massive spin-2}
Equation of motion
\begin{equation}
( \mathfrak{D}^2 -2) v_{(mp)} + 2i \Big( \varepsilon_m^{\;\;n} v_{(np)} - v_{(mn)}\varepsilon^n_{\;\;p}\Big) = 0.
\end{equation}
Lorenz condition
\begin{equation}
\mathfrak{D}^m v_{(mn)} = - 2i \tilde{\varepsilon}_{nm} C^m - 2i \tilde{\varepsilon}_{nm} a^m -2i \varepsilon_{nm} \mathfrak{D}^m \phi.
\end{equation}
Other constraints for $v_{mn}$
\begin{align}
&v^n_{\;\; n} + 8 \phi = 0, \\
&v_{[mn]} = 2 \epsilon_{mnpq} \Big( \mathfrak{D}^p C^q +  \mathfrak{D}^p a^q \Big) + 2i \varepsilon_{mn} \phi.
\end{align}

\newsec{Conclusions}[conclu]

In this work, we first showed how to couple the hybrid formalism for the open superstring to a constant electromagnetic background and derived expressions for the worldsheet variables in terms of the oscillator modes. We then computed the open superstring field theory action and equations of motion  for the compactification-independent massless sector of the charged open superstring in a constant background, and demonstrated that the result is consistent with previous calculations from bosonic string theory. Finally, we constructed the open superstring field theory action for the first massive compactification-independent states of the charged open superstring in a constant background and showed that the equations of motion in Lorenz gauge describe 12 complex bosonic and 12 complex fermionic propagating degrees of freedom. 

The most obvious application of our results is to try to relate our action with alternative actions for a charged massive spin-$3/2$ field in four dimensions. Although we have shown that the massive spin-$3/2$ field in open superstring field theory propagates causally, it is coupled in four dimensions through the electromagnetic background with a massive spin-$1/2$ field. It is unclear if our action can be generalized in a manner to decouple these massive spin-$3/2$ and spin-$1/2$ fields, except in $d=10$ where there is no massive spin-$1/2$ field in the spectrum of the uncompactified superstring. 

To compute the equations of motion of the massive component fields, we used Lorenz gauge to simplify the computations. In principle, one can explicitly expand the action of \actionmassiveback  in terms of the component fields and obtain a gauge-invariant description of the equations of motion. Although it might be complicated, this gauge-invariant component action could be useful for looking for generalizations which describe the massive spin-$2$ and massive spin-$3/2$ fields without coupling to massive fields of lower spin. For example, in $d=10$, one expects to be able to describe both the spin-$3/2$ and spin-$2$ fields without coupling to fields of lower spin. It would be interesting to see how this is accomplished, and to compare with the bosonic string field theory action of Argyres-Nappi \cite{Argyres:1989cu} in which the massive spin-$2$ field is described in $d=26$ without massive fields of lower spin.

\section*{\centering Acknowledgements}

KB would like to thank ICTP-SAIFR for hospitality at the start of this project. His work is supported in part by the Agence Nationale de Recherche under grant ANR-15-CE31-0002 ``HiggsAutomator''. NB would like to acknowledge partial financial support from CNPq grant number 311434/2020-7 and  FAPESP grant numbers 2016/01343-7, 2019/24277-8 and 2019/21281-4. CAD acknowledges FAPESP grant number 2020/10183-9 for financial support. ML acknowledges FAPESP grant numbers 2019/17805-8 and 2016/16824-0 for financial support. We would also like to thank Massimo Porrati for useful discussions.

\newpage

\begin{appendices}

\newsec{Results in Spinor and Oscillator Algebra}  \label{spinorconventions}

Some relations and definitions about the supersymmetric modes that were used in the calculations of this work are given below.
For a more complete list of identities and conventions for the sigma matrices and spinorial indices, see the appendices of \cite{Wess:1992cp}.

\begin{equation}
(\chi^{\al} \psi_{\al}) = (\chi \psi), \quad (\bar{\chi}_{\ad} \bar{\psi}^{\ad}) = (\bar{\chi} \bar{\psi}), \quad (\chi \psi)^{\dagger} =  (\bar{\chi} \bar{\psi}).
\end{equation}

\begin{equation}
\theta_0^{\al}\theta_{\bt 0}= \frac{1}{2} \delta^{\al}_{\bt}(\theta_0 \theta_0), \quad \bar{\theta}_{\ad 0} \bar{\theta}_0^{\bd}= \frac{1}{2} \delta^{\bd}_{\ad} (\bar{\theta}_0\bar{\theta}_0).
\end{equation}

\begin{align}
&[p_{\al 0}, (\theta_0 \theta_0)] = -2 \theta_{\al 0}, \quad [\bar{p}_{\ad 0}, (\bar{\theta}_0 \bar{\theta}_0)]= -2 \bar{\theta}_{\ad 0}, \\
& [(p_0 p_0),(\theta_0 \theta_0)]=-4 -4(\theta_0 p_0), \quad  [(\bar{p}_0 \bar{p}_0), (\bar{\theta}_0 \bar{\theta}_0)] = -4 -4 (\bar{\theta}_0 \bar{p}_0).
\end{align}

\begin{align}
\al_{0n} \al_{0p} (\sigmabar^n \sigma^p)^{\ad}_{\;\; \bd} & = -\al_{0n} \al^n_0 \delta^{\ad}_{\bd} - i \varepsilon_{np} (\sigmabar^{np})^{\ad}_{\;\;\bd}, \\
\al_{0n} \al_{0p} (\sigma^n \sigmabar^p)_{\al}^{\;\; \bt} & = -\al_{0n} \al^n_0 \delta_{\al}^{\bt} - i \varepsilon_{np} (\sigma^{np})_{\al}^{\;\; \bt}.
\end{align}

\begin{equation}
[d^{\al}_n,\Pi_{\al \ad m}] = 8i \pd \bar{\theta}_{\ad m+n}, \quad [\dbar^{\ad}_n, \Pi_{\al \ad m}]= -8i \pd \theta_{\al m+n}.
\end{equation}

\begin{equation}
\begin{split}
[d_{\al m}, \dbar_{\ad n} \dbar^{\ad}_n] &= 4i \Pi_{\al \ad m+n} \dbar^{\ad}_n + 16 \pd \theta_{\al m+2n} \\
& =4i \dbar^{\ad}_n \Pi_{\al \ad m+n} - 16 \pd \theta_{\al m+2n}
\end{split}
 \quad 
 \begin{split}
 [\dbar_{\ad m}, d^{\al}_n d_{\al n}]&= -4i \Pi_{\al \ad m+ n} d^{\al}_n + 16 \pd \bar{\theta}_{\ad m+2n} \\ 
 &= - 4i d^{\al}_n \Pi_{\al \ad m+n} - 16 \pd \bar{\theta}_{\ad m+2n}.
 \end{split}
\end{equation}

\begin{equation}
d^{\al}_0 \Pi_{\al \ad 0} \dbar^{\ad}_0 = -\dbar^{\ad}_0 \Pi_{\al \ad 0} d^{\al}_0 -4i\Pi^n_0 \Pi_{n 0} +8i \pd \theta^{\al}_0 d_{\al 0} + 8i \pd \bar{\theta}_{\ad 0} \dbar^{\ad}_0.
\end{equation}

\begin{equation}
\begin{split}
[d^{\al}_0 d_{\al 0}, \dbar_{\ad 0} \dbar^{\ad}_0] &= -8i\dbar^{\ad}_0 \Pi_{\al \ad 0} d^{\al}_0 + 16 \Pi^n_0 \Pi_{n 0} -32 \pd \bar{\theta}_{\ad 0} \dbar^{\ad}_0 -32 \pd \theta^{\al}_0 d_{\al 0} \\
& = 8i d^{\al}_0 \Pi_{\al \ad 0} \dbar^{\ad}_0 -16 \Pi^n_0 \Pi_{n0} + 32 \pd \bar{\theta}_{\ad 0} \dbar^{\ad}_0 + 32 \pd \theta^{\al}_0 d_{\al 0} \\
& = 4i \Pi_{\al \ad 0} [d^{\al}_0, \dbar^{\ad}_0] + 32 d^{\al}_0 \pd \theta_{\al 0} -32 \pd \bar{\theta}_{\ad 0} \dbar^{\ad}_0.
\end{split}
\end{equation}

\begin{align}
&d_{\al 0} f(\theta_0, \thetab_0)= \big[- p_{\al 0} - (\sigma^m \bar{\theta}_0)_{\al} \al_{0m} - \frac{i}{2} (\bar{\theta}_0 \bar{\theta}_0) (\varepsilon \cdot \sigma)_{\al}^{\; \bt} \theta_{\bt 0}\big] f(\theta_0, \thetab_0), \\
&\dbar_{ \ad 0} f(\theta_0, \thetab_0)= \big[ -\bar{p}_{\ad 0} + (\theta_0 \sigma^m)_{\ad} \al_{0m} + \frac{i}{2} (\theta_0 \theta_0) (\varepsilon \cdot \sigmabar)^{\bd}_{\; \ad} \bar{\theta}_{\bd 0}\big]f(\theta_0, \thetab_0), \\
&\Pi^m_0f(\theta_0, \thetab_0) = \big[- i \al_0^m + \frac{i}{2} \epsilon^{mrsn} \varepsilon_{rs} (\theta_0 \sigma_n \bar{\theta}_0)\big]f(\theta_0, \thetab_0).
\end{align}

\begin{align}
&d^{\al}_0 d_{\al 0} f(\theta_0, \thetab_0)=\big[ p_0^2 -2 (\bar{\theta}_0 \sigmabar^m)^{\al} \al_{0m} p_{\al 0} + (\bar{\theta}_0 \bar{\theta}_0) \al^n_0 \al_{0n} -i (\bar{\theta}_0 \bar{\theta}_0) (\varepsilon \cdot \sigma)_{\bt}^{\;\al} \theta_0^{\bt} p_{\al 0}\big]f(\theta_0, \thetab_0), \\
&\dbar_{\ad 0}\dbar^{\ad}_0 f(\theta_0, \thetab_0)= \big[\bar{p}_0^2 -2 (\sigmabar^m \theta_0)^{\ad} \al_{0m} \bar{p}_{\ad 0} + (\theta_0 \theta_0) \al^n_0 \al_{0n} - i (\theta_0 \theta_0) (\varepsilon \cdot \sigmabar)^{\bd}_{\;\ad} \bar{\theta}_{\bd 0} \bar{p}_0^{\ad}\big]f(\theta_0, \thetab_0), \\
& \Pi^m_0 \Pi_{m0} f(\theta_0, \thetab_0)=\big[- \al^m_0 \al_{0m} + \al_{0m} \epsilon^{mrsn} \varepsilon_{rs} (\theta_0 \sigma_n \bar{\theta}_0) - \frac{1}{2} \varepsilon_{rs} \varepsilon^{rs} (\theta_0 \theta_0) (\bar{\theta}_0 \bar{\theta}_0)\big]f(\theta_0, \thetab_0).
\end{align}

\begin{equation}
\{ d_{\al 0}, \pd \theta^{\bt}_0 \} = \frac{i}{2} (\varepsilon \cdot \sigma)_{\al}^{\; \bt}, \quad \{ \dbar^{\ad}_0, \pd \thetab_{\bd 0} \} = \frac{i}{2} (\varepsilon \cdot \sigmabar)^{\ad}_{\;\bd}.
\end{equation}

\begin{equation}
[d_0^2, \pd \theta^{\bt}_0] = i (\varepsilon \cdot \sigma)_{\al}^{\;\bt} d_0^{\al}, \quad [\dbar_0^2, \pd \thetab_{\bd 0}] = i (\varepsilon \cdot \sigmabar)^{\ad}_{\; \bd} \dbar_{\ad 0}.
\end{equation}

\begin{equation}
\label{Deltas}
\Delta_{\al}^{\; \bt}  = \delta^{\bt}_{\al} + \frac{i}{2}  (\varepsilon \cdot \sigma)^{\;\bt}_{\al}, \quad \bar{\Delta}^{\ad}_{\;\bd}= \delta^{\ad}_{\bd} + \frac{i}{2}  (\varepsilon \cdot \sigmabar)^{\ad}_{\;\bd}.
\end{equation}

\begin{align}
&(-d_0^2 \dbar_1^2 - d_1^2 \dbar_0^2 + d_2^2 \dbar_{-1}^2) \varphi = (-\dbar_1^2 d_0^2 + \dbar_2^2 d_{-1}^2 - \dbar_0^2 d_1^2) \varphi,  \\
& (3 d_0^2 \dbar_1^2 -3 d_1^2 \dbar_0^2 + d_2^2 \dbar_{-1}^2) \varphi = (-3 \dbar_0^2 d_1^2 + 3 \dbar_1^2 d_0^2 - \dbar_2^2 d_{-1}^2) \varphi,  \\
& (d_{-1}^2 \dbar_1^2 -2 d_0^2 \dbar_0^2 + d_1^2 \dbar_{-1}^2) \varphi = (\dbar_{-1}^2 d_1^2 + \dbar_1^2 d_{-1}^2 -2 \dbar_0^2 d_0^2) \varphi, \label{relationsforeommodes}  \\
& \dbar_1^2 d_{-1}^2 B = d_1^2 \dbar_{-1}^2 B -2 [d_0^2, \dbar_0^2]B, \\
& (64i d^{\al}_0 \Pi_{\al \ad 0} \dbar^{\ad}_0 -19 d_0^2 \dbar_0^2 -3 \dbar_0^2 d_0^2)C =\big[-11\{d_0^2, \dbar_0^2 \} + 128 \Pi^n_0 \Pi_{n0} -256 (\pd \bar{\theta}_0 \dbar_0) \nonumber \\ 
& \hspace{6.11cm}-256 ( \pd \theta_0 d_0) \big]C. 
\end{align}

\begin{align}
d_0^2 \dbar_1^2 \varphi &= 2(d_0 d_0) (\dbar_0 \dbar_1) \varphi, \\
d_1^2 \dbar_0^2 \varphi &= (8i d^\al_0 \Pi_{\al \ad 0} \dbar_1^{\ad} + 2 d^{\al}_0 (\dbar_0 \dbar_0) d_{\al 1} + 8i d^{\al}_0 \dbar^{\ad}_0 \Pi_{\al \ad 1}  -32 d^{\al}_0 \pd \theta_{\al 1} ) \varphi, \\
d_2^2 \dbar_{-1}^2 \varphi &= (-16 \Pi_{\al \ad 0} \Pi^{\ad \al}_1 - 8i\Pi_{\al \ad 0} \dbar^{\ad}_0 d^{\al}_1 + 8i d^{\al}_0 \Pi_{\al \ad 0} \dbar^{\ad}_1 + 8i d^{\al}_0 \dbar^{\ad}_0 \Pi_{\al \ad 1} \nonumber\\
&- 32 \pd \bar{\theta}_{\ad 0} \dbar^{\ad}_1 -32 \dbar_{\ad 0} \pd \bar{\theta}^{\ad}_1 -64 d^{\al}_0 \pd \theta_{\al 1} )\varphi,  \\
 d^2_{-1} \dbar_1^2 \varphi &= 4(d_{-1} d_0)(\dbar_0 \dbar_1) \varphi, \\
 d_1^2 \dbar_{-1}^2 \varphi &= ( 8i d^{\al}_0 \Pi_{\al \ad 0} \dbar^{\ad}_0 + 8i \Pi_{\al \ad -1} d^{\al}_0 \dbar^{\ad}_1 - 16 \Pi_{\al \ad -1} \Pi^{\ad \al}_1-8i \Pi_{\al \ad -1} \dbar^{\ad}_0 d^{\al}_1\nonumber \\
 & -8i \dbar^{\ad}_{-1} d^{\al}_0 \Pi_{\al \ad 1} -4 \dbar_{\ad -1} d^{\al}_0 \dbar^{\ad}_0 d_{\al 1} + 8i d^{\al}_{-1} \Pi_{\al \ad 0} \dbar^{\ad}_1 + 8i d^{\al}_{-1} \dbar^{\ad}_0 \Pi_{\al \ad 1}  \nonumber \\
 & -64 d^{\al}_{-1} \pd \theta_{\al 1} -64 \pd \bar{\theta}_{\ad -1} \dbar^{\ad}_1) \varphi,   \\
 d_0^2 \dbar_0^2 \varphi &= ( d^{\al}_0 d_{\al 0} \dbar_{\ad 0} \dbar^{\ad}_0 + 2 \dbar_{\ad -1} d^{\al}_0 d_{\al 0} \dbar^{\ad}_1 + 2 d^{\al}_{-1} \dbar_{\ad 0} \dbar^{\ad}_0 d_{\al 1} + 8i d^{\al}_{-1} \Pi_{\al \ad 0} \dbar^{\ad}_1 \nonumber\\
 & + 8i d^{\al}_{-1} \dbar^{\ad}_0 \Pi_{\al \ad 1} + 8i \Pi_{\al \ad -1} d^{\al}_0 \dbar^{\ad}_1 -32 \pd \bar{\theta}_{\ad -1} \dbar^{\ad}_1 -32 d^{\al}_{-1} \pd \theta_{\al 1}) \varphi. 
\end{align}

\begin{align}
d_0^2 \dbar_0^2 B &= (d_0 d_0) (\dbar_0 \dbar_0) B,\\
d_1^2 \dbar_{-1}^2 B &= 8i d^{\al}_0 \Pi_{\al \ad 0} \dbar^{\ad}_0 B, \\
d_{-1}^2 \dbar_0^2 B & = 2 d^{\al}_{-1} d_{\al 0} \dbar_{\ad 0} \dbar^{\ad}_0 B, \\
d_0^{2} \dbar^2_{-1} B & = ( 2 \dbar_{\ad -1} d^{\al}_0 d_{\al 0} \dbar^{\ad}_0 + 8i d^{\al}_{-1} \Pi_{\al \ad 0} \dbar^{\ad}_0 + 8i \Pi_{\al \ad -1} d^{\al}_0 \dbar^{\ad}_0 -32 \pd \bar{\theta}_{\ad -1} \dbar^{\ad}_0) B, \\
\dbar^2_{-1} d^2_0 B &= 2 \dbar_{\ad -1} \dbar^{\ad}_0 d^{\al}_0 d_{\al 0} B, \\
\dbar^2_0 d^2_{-1} B & = ( 2 d^{\al}_{-1} \dbar_{\ad 0} \dbar^{\ad}_0 d_{\al 0} + 8i \dbar_{\ad -1} \Pi^{\ad \al}_0 d_{\al 0} + 8i \Pi_{\al \ad -1} \dbar^{\ad}_0 d^{\al}_0 -32 \pd \theta^{\al}_{-1} d_{\al 0} ) B.
\end{align}

\end{appendices}

\newpage

\bibliography{INSPIRE}

\end{document}